\documentclass[onecolumn,aps,prd,showpacs,floatfix]{revtex4}

\usepackage{seceqn}
\usepackage{amsmath}
\usepackage{amsfonts}
\usepackage{amssymb}
\usepackage{bm}
\usepackage{bbm}
\usepackage{graphicx}
\usepackage{color}
\usepackage{cancel}
\usepackage{maybemath}
\usepackage{fancybox}

\newcommand{\redcol}{black}
\newcommand{\bluecol}{black}

\newcommand{\dd}{\mathrm{d}}
\newcommand{\ee}{\mathrm{e}}
\newcommand{\ii}{\mathrm{i}}

\newcommand{\calC}{\mathcal{C}}
\newcommand{\calJ}{\mathcal{J}}
\newcommand{\calL}{\mathcal{L}}

\newcommand{\calP}{\mathcal{P}}
\newcommand{\calT}{\mathcal{T}}
\newcommand{\calU}{\mathcal{U}}
\newcommand{\calV}{\mathcal{V}}

\newcommand{\plus}{{\mbox{{\bf{\tiny +}}}}}

\newcommand{\trc}{\mathrm{tr}}

\newcommand{\eV}{\mathrm{eV}}

\newcommand{\GeV}{\mathrm{GeV}}

\begin{document}

\title{From Generalized Dirac Equations to a Candidate for Dark Energy}

\author{U. D. Jentschura}
\author{B. J. Wundt}
\affiliation{Department of Physics,
Missouri University of Science and Technology,
Rolla, Missouri 65409, USA}

\begin{abstract}
We consider extensions of the Dirac equation with mass terms $m_1 + \ii \,
\gamma^5 m_2$ and $\ii \, m_1 + \gamma^5 \, m_2$. The corresponding
Hamiltonians are Hermitian and pseudo-Hermitian (``$\gamma^5$ Hermitian''),
respectively. The fundamental spinor solutions for all generalized 
Dirac equations are found in the helicity basis
and brought into concise analytic form.  We postulate that the time-ordered product of
field operators should yield the Feynman propagator ($\ii \, \epsilon$
prescription), and we also postulate that the tardyonic as well as tachyonic
Dirac equations should have a smooth massless limit.  These postulates lead to
sum rules that connect the form of the fundamental field anticommutators with
the tensor sums of the fundamental plane-wave eigenspinors and the projectors
over positive-energy and negative-energy states. In the massless case, the sum
rules are fulfilled by two egregiously simple, distinguished functional forms.
The first sum rule remains valid in the case of a tardyonic theory and leads to
the canonical massive Dirac field. The second sum rule is valid for a
tachyonic mass term and 
leads to a natural suppression of the right-handed helicity states for
tachyonic particles, and left-handed helicity states for tachyonic spin-$1/2$
antiparticles.  When applied to neutrinos, the theory contains a free tachyonic
mass parameter. Tachyons are known to be repulsed by gravity. 
We discuss a
possible role of a tachyonic neutrino as a contribution to the accelerated
expansion of the Universe (``dark energy'').
\end{abstract}


\maketitle 

\tableofcontents

\newpage

%
%
\section{Introduction}
\label{intro}

%
%
\subsection{Generalized Dirac Equations: Mass Terms and Dispersion Relations}

Dirac is often quoted as saying in some of his talks that the equation that
carries his name~\cite{Di1928a,Di1928b} is ``more intelligent than its inventor''.
Of course, it needs to be added that it was Dirac himself who found most of the
additional insight.  
Here, we are concerned with of extensions of the Dirac
equation which contain both tardyonic and tachyonic mass terms. 
Tardyonic (subluminal) mass
terms lead to dispersion relations of the form $E = \sqrt{\vec p^{\,2} + m^2}$,
whereas tachyonic mass terms lead to superluminal dispersion relations of the
form $E = \sqrt{\vec p^{\,2} - m^2}$, where $E$ is the
energy and $\vec p$ is the momentum. 
The generalized, matrix-valued mass term $M$ enters the Dirac 
equation in the form $(\ii \, \gamma^\mu \partial_\mu - M) \, \psi(x) = 0$. 
The $\gamma^\mu$ are $4 \times 4$ matrices
that fulfill the relations 
$\{ \gamma^\mu, \gamma^\nu \} = 2 g^{\mu\nu}$ where we choose the 
space-time metric as $g^{\mu\nu} = \mathrm{diag}(1,-1,-1,-1)$.
The $\partial_\mu$ denote  the partial derivative $\partial/\partial x^\mu$
with respect to the space-time coordinate $x^\mu = (t, \vec x)$.
It is quite surprising that a systematic presentation of the
solutions of the generalized Dirac equations
$(\ii \, \gamma^\mu \partial_\mu - M) \, \psi(x) = 0$,
in the helicity basis~\cite{JaWi1959}, has not been 
recorded in the literature to the best of our knowledge.
While the following discussion is somewhat technical,
we believe that it will be beneficial to give their explicit
form, in order to fix ideas for the following discussion.

We set $\hbar = c = \epsilon_0 = 1$ and
use the Dirac matrices in the standard representation
\begin{equation}
\gamma^0 =\beta = \left( \begin{array}{cc} \mathbbm{1}_{2\times 2} & 0 \\
0 & -\mathbbm{1}_{2\times 2} \\
\end{array} \right) \,,
\qquad
\vec\gamma = \left( \begin{array}{cc} 0 & \vec\sigma \\ -\vec\sigma & 0  \\
\end{array} \right) \,,
\qquad
\gamma^5 = \left( \begin{array}{cc} 0 & \mathbbm{1}_{2\times 2} \\
\mathbbm{1}_{2\times 2} & 0  \\
\end{array} \right) \,,
\end{equation}
and define $\vec\alpha = \gamma^0 \vec\gamma$.
For the ordinary Dirac theory, one has
$M = m_1$ (one should say more precisely $M = m_1 \, \mathbbm{1}_{4 \times 4}$)
with a real mass $m_1$,
\begin{equation}
\label{eq1}
\left( \ii \gamma^\mu \, \partial_\mu - m_1 \right) \, \psi(x) = 0 \,.
\end{equation}
The dispersion relation is $E = \sqrt{\vec p^{\,2} + m_1^2}$.
The corresponding Dirac Hamiltonian reads
\begin{equation}
\label{HD1}
H^{(1)} = \vec \alpha \cdot \vec p + \beta \,  m_1\,.
\end{equation}
Extensions of the Dirac equation with pseudoscalar mass terms
that contain the fifth current have been introduced in the literature.
In Ref.~\cite{JoKaPr2000}, it is shown that for a
mass term of the form $M = m_1 + \ii \gamma^5 m_2$,
the fermion propagator may
obtain nontrivial gradient corrections already at
the first order in derivative expansion,
for a position-dependent mass.
In that case, the fermion self-energy may contribute to a
conceivable explanation
for $\calC \calP$-violation during electroweak
baryogenesis, as pointed out in Ref.~\cite{JoKaPr2000}.
We thus study the following, generalized form of the 
tardyonic (subluminal) Dirac equation,
\begin{equation}
\label{eqt}
\left( \ii \gamma^\mu \, \partial_\mu - m_1 - \ii \, \gamma^5 \, m_2
\right) \, \psi(x) = 0 \,.
\end{equation}
The dispersion relation is $E = \sqrt{\vec p^{\,2} + m_1^2 + m_2^2}$.
The Hermitian tardyonic Hamiltonian operator reads as
\begin{equation}
\label{HDt}
H^{(t)} = \vec \alpha \cdot \vec p + \beta \, m_1 +
\ii \, \beta \, \gamma^5 \, m_2 \,.
\end{equation}
We may indicate a further motivation for our study; namely,
the unitarity of the $S$ matrix
implies the existence of useful relations~\cite{Ma2012priv}
for the even powers $(m_2)^{2n}$ obtained
upon expanding a one-loop amplitude,
formulated with a mass term $m_1 + \ii \, \gamma^5 \, m_2$,
in powers of $m_2$. This implies that a better understanding of the
tardyonic equation with two mass terms
could be of much more general interest.

It has not escaped our attention that the chiral transformation 
\begin{equation}
\mu \exp(\ii \, \gamma^5 \, \theta) = 
\mu \cos \, \theta + \ii \, \gamma^5 \, \mu \sin \, \theta = 
m_1 + \ii \, \gamma^5 \, m_2
\end{equation}
connects the two Hamiltonians $H^{(1)}$ and $H^{(t)}$
for $m_1 = \mu \, \cos\,\theta$ and $m_2 = \mu \, \sin\,\theta$,
but it is computationally easier and more instructive 
to consider the real and imaginary parts of the mass term separately.

Within a systematic
approach to generalized Dirac equations with pseudoscalar mass 
terms, we also consider tachyonic (superluminal)
mass terms of the form $M = \gamma^5 \, m$ which induce a 
superluminal dispersion relation $E = \sqrt{\vec p^{\,2} - m^2}$.
The corresponding generalized Dirac equation has been named the 
``tachyonic Dirac equation'' and reads as 
follow~\cite{ChHaKo1985,Ch2000,Ch2002,JeWu2012epjc,JeWu2012jpa},
\begin{equation}
\label{eq5}
\left( \ii \gamma^\mu \, \partial_\mu - \gamma^5 \, m \right) \, 
\psi(x) = 0  \,.
\end{equation}
The corresponding Hamiltonian reads
\begin{equation}
\label{HD5}
H_5 = \vec \alpha \cdot \vec p + \beta \, \gamma^5 \, m \,.
\end{equation}
The relation 
$H_5 = \gamma^0 \, H_5^+(-\vec r) \, \gamma^0$
has been given in~Refs.~\cite{JeWu2012epjc,JeWu2012jpa}. However, 
it is much more instructive to observe that 
$H_5$ is $\gamma^5$ Hermitian, i.e., $H_5 = \gamma^5 H_5^+ \gamma^5$.
The concept of $\gamma^5$ Hermiticity is known in 
lattice theory~\cite{AkDaSpVe2010,GaLa2009}
and is otherwise called pseudo-Hermiticity~\cite{Pa1943,BeBo1998,BeDu1999,%
BeBoMe1999,BeWe2001,BeBrJo2002,
Mo2002i,Mo2002ii,Mo2003npb,JeSuZJ2009prl,JeSuZJ2010}.

An obvious generalization of the tachyonic case contains an imaginary mass and 
a $\gamma^5$ mass term,
\begin{equation}
\label{eqp}
\left( \ii \gamma^\mu \, \partial_\mu - 
\ii \, m_1 - \gamma^5 m_2 \right) \, \psi(x) = 0 \,.
\end{equation}
The dispersion relation is $E = \sqrt{\vec p^{\,2} - m_1^2 - m_2^2}$.
The corresponding Hamiltonian reads
\begin{equation}
\label{HDp}
H' = \vec \alpha \cdot \vec p + \ii \, \beta \,  m_1 + \beta \, \gamma^5 \, m_2\,,
\end{equation}
and is $\gamma^5$ Hermitian,
$H' = \gamma^5 H'^\plus \gamma^5$.
For $m_2 = 0$, Eq.~\eqref{eqp} has been discussed in 
Refs.~\cite{BaSh1974,Je2012imag}.

It is our goal here to present the fundamental eigenspinors corresponding to
the plane-wave solution of the equations~\eqref{eq1},~\eqref{eqt},~\eqref{eq5},
and~\eqref{eqp} in a unified and systematic manner. Furthermore, we discuss the
second-quantized versions of the fermionic theories described by the
generalized Dirac equations. Anticipating part of the results, we may point 
out that the massless Dirac equation ``interpolates'' between the tardyonic
equations~\eqref{eq1} and~\eqref{eqt} and the tachyonic equations~\eqref{eq5}
and~\eqref{eqp}. For zero mass, helicity and chirality
are equal. Helicity and chirality ``depart'' from each other
in very specific directions, when the tardyonic and tachyonic mass terms
are ``switched on'', as we shall discuss in the following.

%
%
\subsection{Tachyonic Dirac Equation and Neutrinos: Possible Connections}

The tachyonic generalized Dirac equations~\eqref{eq5} and~\eqref{eqp}
describe the motion of superluminal particles,
which may either be important for astrophysical studies (neutrinos)
or for artificially generated environments such as honeycomb photonic
lattices in which pertinent dispersion relations become practically 
important~\cite{SzReBHMo2012}.
Contrary to other, somewhat catchy, claims, the existence of superluminal
particles would {\em not} falsify Einstein's theory of
special relativity~\cite{Sz2012}, which according to common wisdom is based on the
following postulates:
(i)~The principle of relativity states that the
laws of physics are the same for all observers in uniform motion relative
to one another.
(ii)~The speed of light in a vacuum is the same for all observers, regardless of
their relative motion or of the motion of the source of the light.
Predictions of relativity theory regarding the relativity of simultaneity,
time dilation and length contraction
would not change if superluminal particles did exist.
Furthermore, as shown by Sudarshan {\em et al.}
(Refs.~\cite{BiDeSu1962,ArSu1968,DhSu1968,BiSu1969}) and Feinberg
(Refs.~\cite{Fe1967,Fe1977}),
the existence of tachyons, which are superluminal particles
fulfilling a Lorentz-invariant dispersion relation
$E^2 = \vec p^2 - m_\nu^2$, is fully compatible with
special relativity and Lorentz invariance.
According to special relativity, it is forbidden to
accelerate a particle ``through'' the light barrier
(because $E = m/\sqrt{1- v^2} \to \infty$ for $v \to 1$),
but a genuinely superluminal particle remains superluminal
upon Lorentz transformation.
Significant problems are encountered when one attempts
to quantize the tachyonic theories, but again, as
shown in Ref.~\cite{JeWu2012epjc}, these problems may not be as serious
as previously thought. In particular, the so-called reinterpretation
of solutions propagating into the past according to the
Feynman prescription~\cite{BiSu1969} is a cornerstone of modern field theory.
Furthermore, it has been shown in Ref.~\cite{JeWu2012epjc}
that tachyonic particles can be localized, and
equal-time anticommutators of the spin-$1/2$ tachyonic field
involve an unfiltered Dirac-$\delta$
[see Eq.~(37) of Ref.~\cite{JeWu2012epjc}].

Despite these arguments, we can say that,
from the point of view of fundamental symmetries,
accepting a superluminal neutrino would be equivalent to 
an ``an ugly duckling''. Adding to the difficulties,
we notice that recent experimental claims 
regarding the conceivable observation of highly 
superluminal neutrinos have turned out to be false.
One may point out that a relative deviation $v = c\,(1+\delta)$ with 
$\delta \sim 10^{-5}$ at $E \sim 30 \,{\rm GeV}$,
as claimed by some recent experimental collaborations,
would correspond to a negative 
neutrino mass square in the order of $\sim -(100\,{\rm MeV})^2$,
if one assumes a Lorentz-invariant dispersion relation~\cite{Je2012cejp}.
Still, there is at present no conclusive answer regarding 
the conceivable superluminality of at least one
neutrino flavor~\cite{Eh1999a,Eh1999b,Eh2000,Eh2011neutrino,Eh2012neutrino}, and it is intriguing that 
all available direct measurements of the neutrino mass 
square have resulted in negative expectation values,
still compatible with zero within experimental uncertainty,
whereas published experimental best estimates
for the neutrino speed~\cite{KaEtAl1979,AdEtAl2007,OPERA2011v4,ICARUS2012}
have been superluminal, again still compatible with the speed of light 
within experimental error.
The recent ICARUS result~\cite{ICARUS2012} is consistent with this
trend~\cite{RoEtAl1991,AsEtAl1994,StDe1995,AsEtAl1996,%
WeEtAl1999,LoEtAl1999,BeEtAl2008}; the best estimate for the
neutrino velocity is superluminal, but the
deviation from $v_\nu=c$ is statistically insignificant.
The OPERA collaboration~\cite{OPERA2011dracos}  has indicated a preliminary, revised
result of $(v-c)/c = (2.7 \pm 3.1 \mbox{(stat)} {}^{+3.8}_{-2.8} \mbox{(sys.)})
\times 10^{-6}$. Neither subluminal nor superluminal propagation
velocities are excluded based on the available experimental data.
The ``ugly duckling of a superluminal neutrino'' is not beautiful;
if we are to consider accepting it,
then we should be able to hope that the emergence of at least
one ``swan'' (or ``intellectual benefit'') 
should be the result of this operation.

Before we discuss the possible emergence of these benefits, let us include
some historic remarks.  According to reliable sources~\cite{Fi2012a}, 
Professor J.~A.~Wheeler,
in his later years at the University of Austin (Texas), used to argue that the
neutrino has to be massless, necessarily, and that in his opinion, it could only
be a massless Weyl particle with definite helicity (and chirality).  Recently
presented arguments~\cite{Je2012nemeti} regarding the possibility of overtaking a
subluminal, left-handed neutrino,
looking back and seeing a right-handed
neutrino, were supposedly already used by Wheeler in order to dispel the
conceivable existence of a neutrino mass term.  This paradox has been termed
``autobahn paradox'' in Ref.~\cite{Je2012nemeti} and excludes a Dirac neutrino unless one
assumes exotic processes like sterile right-handed massive neutrinos.  [The
problem with a right-handed sterile massive neutrino is that for massive
neutrinos, chirality and helicity are different, hence a $V-A$ coupling of the
form $\gamma^\mu (1-\gamma^5)$ no longer vanishes for massive Dirac neutrinos
if one uses the canonical eigenstates of the massive Dirac equation.
One therefore has to invoke additional exotic mechanisms in order to 
ensure the ``sterility'' of the right-handed Dirac neutrinos.] Wheeler
also disliked~\cite{Fi2012a} the notion of a Majorana neutrino, arguing that the
charge conjugation invariance condition imposed on the Majorana particle
precludes the existence of plane-wave solutions to the Majorana equation,
and maximally violates lepton number.
Again, these arguments~\cite{Fi2012b} are in full agreement with those
recently given in Ref.~\cite{Je2012nemeti}.

The original standard model thus called for 
manifestly massless neutrinos.
The commonly accepted observation of neutrino oscillations
precludes the possibility that all three 
generations of neutrino mass eigenstates are massless.
Lepton number conservation is based on the global gauge 
symmetry $\psi \to \psi \, \exp(\ii \Lambda)$,
applied simultaneously to all
lepton fields. A Majorana neutrino would destroy lepton number 
as a global symmetry but 
solve the ``autobahn paradox'', because a Majorana neutrino 
would be equal to its own antiparticle and thus, looking back,
the right-handed neutrino state would consist of the same
particle$=$antiparticle.

On the other hand, if we assume that the neutrino is described by the tachyonic 
Dirac equation, then the following statements are valid:
\begin{itemize}
\item
Statement \#{}1: We can properly assign lepton number and 
use plane-wave eigenstates for incoming and outgoing 
particles, while allowing for nonvanishing mass
terms and thus, mass square differences among the 
neutrino mass (not flavor) eigenstates.
\item
Statement \#{}2: There is a 
natural resolution for the ``autobahn paradox''
because a left-handed spacelike neutrino always remains
spacelike upon Lorentz transformation and cannot be overtaken.
\item
Statement \#{}3: The right-handed particle and left-handed 
antiparticle states are suppressed due to negative 
Fock-space norm.
\item
Statement \#{}4: At least qualitatively, tachyonic neutrinos
could yield an explanation for a repulsive force on 
intergalactic distance scales as they are repulsed,
like all tachyons, by gravitational interactions (``dark energy'').
\end{itemize}
Pauli~\cite{Pa1930} postulated the existence of neutrinos, on the basis
of the conservation of angular momentum and energy, and also introduced
pseudo-Hermitian operators~\cite{Pa1943}. 
Here, we describe conceivable connections of neutrino physics and
pseudo-Hermitian operators.
Final clarification can only come from experiment.
When in 1956, F.~Reines and C.~Cowan~\cite{ReCo1956}
discovered the electron neutrino,
two and a half years before Pauli's death, Pauli replied~\cite{EnvM1994} by
telegram: ``Thanks for message. Everything comes to him who knows
how to wait. Pauli.'' In defense of the tachyonic hypothesis,
we would like to stress that a tachyonic Dirac neutrino 
would allow us to retain lepton number conservation 
as a symmetry of nature.
We would thus like to write up these thoughts in the 
current article, with attention to detail.
We should point out that our approach fully conserves Lorentz invariance, 
in contrast to the extensions of the Standard Model 
based on Lorentz-violating terms which can otherwise
lead to superluminal propagation
(see Tables~11 and~13 of Ref.~\cite{KoMe2012}).

Units with $\hbar = c = \epsilon_0 = 1$ are used
throughout the paper.  The organization is as follows: In Sec.~\ref{sec2}, we
discuss massless and tardyonic theories.  Tachyonic extensions of the Dirac
equation are discussed in Sec.~\ref{sec3}.  Connections of the fundamental
tensor sums over the eigenspinors with the derivation of the time-ordered
propagator are analyzed in Sec.~\ref{sec4}. 
A candidate for dark energy is presented in Sec.~\ref{sec5}.
Conclusions are reserved for Sec.~\ref{sec6}. 

%
%
\section{Generalized Dirac Equations: Massless and Tardyonic Theories}
\label{sec2}

%
%
\subsection{Massless Dirac Theory}
\label{sec21}

The massless Dirac equation and the massless
Dirac Hamiltonian read as
\begin{equation}
\ii \gamma^\mu \,\partial_\mu \; \psi(x) = 0 \,,
\qquad
H_0 = \vec\alpha \cdot \vec p \,.
\end{equation}
We note that $H_0$ is both Hermitian as well as 
$\gamma^5$ Hermitian, i.e., $H_0 = \gamma^5 \, H_0^\plus \, \gamma^5$.
The dispersion relation is $E = |\vec k|$.
With $k^\mu = (E, \vec k)$, we seek 
positive-energy and negative-energy solutions 
of the form
\begin{equation}
\psi(x) = u_{\sigma}(\vec k) \, \exp(-\ii k \cdot x) \,, 
\;\;
\phi(x) = v_{\sigma}(\vec k) \, \exp(\ii k \cdot x) \,,
\end{equation}
where $\sigma = \pm$ denotes a quantum number which is 
equal to the helicity for positive-energy states, and 
equal to the negative of the helicity for negative-energy states.
With $\cancel{k} = \gamma^\mu \, k_\mu$, 
we have $\cancel{k} \, u_\pm(k)  = \cancel{k} \, v_\pm(k) = 0$.
In the massless limit, the solutions to the Dirac equation are
given as (see Chap.~2 of Ref.~\cite{ItZu1980})
\begin{subequations}
\label{uv}
\begin{align}
u_+(\vec k) = & \;
\frac{1}{\sqrt{2}} 
\left( \begin{array}{c}
a_+(\vec k) \\[0.33ex]
a_+(\vec k) \\
\end{array} \right) \,,
\;\;
u_-(\vec k) = 
\frac{1}{\sqrt{2}} 
\left( \begin{array}{c}
a_-(\vec k) \\[0.33ex]
-a_-(\vec k) \\
\end{array} \right) \,,
\\
v_+(\vec k) = & \; -u_+(\vec k) \,,
\;\;
v_-(\vec k) = -u_-(\vec k) \,.
\end{align}
\end{subequations}
The well-known helicity spinors are recalled as
\begin{equation}
a_+(\vec k) = \left( \begin{array}{c} 
\cos\left(\frac{\theta}{2}\right) \\[0.33ex]
\sin\left(\frac{\theta}{2}\right) \, \ee^{\ii \, \varphi} \\
\end{array} \right) \,,
\quad
a_-(\vec k) = \left( \begin{array}{c} 
-\sin\left(\frac{\theta}{2}\right) \, \ee^{-\ii \, \varphi} \\[0.33ex]
\cos\left(\frac{\theta}{2}\right) \\
\end{array} \right) \,.
\end{equation}
These fulfill the fundamental relations
$(\vec \sigma \cdot \hat{\vec k}) \, a_\sigma(\vec k) = 
\sigma \, a_\sigma(\vec k)$,
as well as
$\sum_\sigma a_\sigma(\vec k) \otimes a^\plus_\sigma(\vec k) = 
\mathbbm{1}_{2 \times 2}$,
and $\sum_\sigma \, \sigma \; a_\sigma(\vec k) \otimes a^\plus_\sigma(\vec k) = 
\vec \sigma \cdot \hat{\vec k}$, where $\hat{\vec k} = \vec k / | \vec k |$
and the sum over $\sigma$ is over the values $\pm 1$.
The sums over the fundamental bispinors 
$u$ and $v$ fulfill the following sum rules,
\color{\redcol}

\begin{align}
\label{sumrule1}
\mbox{Sum rule I:} 
\qquad
\fbox{$\displaystyle{\sum_\sigma  \; 2 \, |\vec k| \, u_\sigma(\vec k) \otimes
\overline u_\sigma(\vec k) = \cancel{k} \,,
\qquad
\sum_\sigma \; 2 \, |\vec k| \, v_\sigma(\vec k) \otimes
\overline v_\sigma(\vec k) = 
\cancel{k}} \,,$} 
\end{align}
\color{black}
as well as 
\color{\bluecol}
\begin{align}
\label{sumrule2}
\mbox{Sum rule II:} 
\qquad
\fbox{$\displaystyle{\sum_\sigma  \; 2 \, |\vec k| \, (-\sigma) \, u_\sigma(\vec k) \otimes
\overline u_\sigma(\vec k) \, \gamma^5 =
\cancel{k} \,,
\qquad
\sum_\sigma \; 2 \, |\vec k| \, (-\sigma) \, v_\sigma(\vec k) \otimes
\overline v_\sigma(\vec k) \, \gamma^5 =
\cancel{k}}\,.$}
\end{align}
\color{black}
Sum rule I~can be obtained by a quick explicit calculation,
and sum rule~II holds because in the massless limit,
helicity equals $\pm$chirality (positive sign for positive 
energy, negative sign for negative energy). 
We denote the Dirac adjoint as 
$\overline u_\sigma(\vec k) = u^\plus _\sigma(\vec k)  \, \gamma^0$.
One can easily check by an
explicit calculation that 
$\overline u_\sigma(\vec k) \gamma^5 =
\left( \gamma^5 \, \gamma^0 \, u_\sigma(\vec k) \right)^\plus =
(-\sigma) \, \overline u_\sigma(\vec k)$
and
$\overline v_\sigma(\vec k) \gamma^5 =
\left( \gamma^5 \, \gamma^0 \, v_\sigma(\vec k) \right)^\plus =
(-\sigma) \, \overline v_\sigma(\vec k)$.
We can thus introduce a factor $(-\sigma)^2 = 1$ 
under the summation over spins in Eq.~\eqref{sumrule1}
and replace one of the factors $(-\sigma)$ by a multiplication
of the Dirac adjoint spinor from the right by the fifth current.
The Lorentz-invariant normalization of the massless solutions
vanishes, i.e., 
$\overline u_\sigma(\vec k)  \, u_\sigma(\vec k) = 
\overline v_\sigma(\vec k)  \, v_\sigma(\vec k) = 0$.

Eigenstates of the massless Hamiltonian 
$H_0 = \vec \alpha \cdot \vec p$
have to be eigenstates of the chirality operator $\gamma^5$
because the chirality commutes with the Hamiltonian,
in the sense that $\left[ \gamma^5, H_0 \right] = 0$.
Furthermore,
\begin{equation}
H_0 = \vec \alpha \cdot \vec p = 
|\vec p| \, \gamma^5 \, 
\left( \frac{\vec \Sigma \cdot \vec p}{|\vec p|} \right) \,,
\end{equation}
where $\vec\Sigma = \gamma^5 \, \vec\alpha$ is the 
vector of $4 \times 4$ spin matrices, 
and the helicity operator is identified 
as $\vec \Sigma \cdot \vec p/|\vec p|$.
Let $\lambda_1$ be the eigenvalue of chirality and 
$\lambda_2$ be the eigenvalue of the helicity operator.
Then, the eigenvalue of the Hamiltonian is 
$E_0 = |\vec p| \, \lambda_1 \, \lambda_2$.
Since $\lambda_1 = \pm 1$ and $\lambda_2 = \pm 1$,
we easily recover the known fact that helicity equals chirality 
for positive energy, whereas the relation is reversed
for negative-energy states (see also Chap.~2.4 of Ref.~\cite{ItZu1980}).
We are aware of the fact that 
the considerations reported in the current
section partly refer to the literature but we 
give them in some detail because they are essential 
for the following considerations.

%
%
\subsection{Massive Dirac Theory}
\label{sec22}

We start from the ordinary Dirac equation given in 
Eq.~\eqref{eq1}, which reads
$\left( \ii \gamma^\mu \, \partial_\mu - m_1 \right) \, \psi(x) = 0$.
In the helicity basis, the fundamental spinor solutions read as
\begin{equation}
\psi(x) = U^{(1)}_{\pm}(\vec k) \, \exp(-\ii k \cdot x) \,, 
\;\;
\phi(x) = V^{(1)}_{\pm}(\vec k) \, \exp(\ii k \cdot x) \,,
\end{equation}
The algebraic relations that have to be fulfilled by
the bispinor amplitudes $U^{(1)}_\pm(\vec k)$ and 
$V^{(1)}_\pm(\vec k)$ read as follows,
\begin{equation}
\label{U1V1}
\left( \cancel{k} - m_1 \right) \, U^{(1)}_{\pm}(\vec k) = 0 \,, 
\;\;
\left( \cancel{k} + m_1 \right) \, V^{(1)}_{\pm}(\vec k) = 0 \,.
\end{equation}
The dispersion relation is $E^{(1)} = \sqrt{\vec k^2 + m_1^2}$.
In the helicity basis, the solutions
of the equation~\eqref{U1V1} with a tardyonic 
$m_1$ mass term are easily written down,
using the identity $(\cancel{k}-m_1)(\cancel{k}+m_1) = 
k^2 - m_1^2 =0$.
With an appropriate normalization factor,
and after some algebraic simplification,
the positive-energy solutions read as follows,
\begin{subequations}
\label{UU1}
\begin{align}
U^{(1)}_+(\vec k) =& \;
\frac{(\cancel{k} + m_1) \, u_+(\vec k)}%
{\sqrt{(E^{(1)} - |\vec k|)^2 + m_1^2}} 
= \left( \begin{array}{c}
\sqrt{\dfrac{E^{(1)} + m_1}{2 \, E^{(1)}}} a_+(\vec k) \\[2.33ex]
\sqrt{\dfrac{E^{(1)} - m_1}{2 \, E^{(1)}}} a_+(\vec k) 
\end{array} \right) \,,
\\[0.33ex]
U^{(1)}_-(\vec k) =& \;
\frac{(\cancel{k} + m_1)\,u_-(\vec k)}%
{\sqrt{(E^{(1)} - |\vec k|)^2 + m_1^2}} 
= \left( \begin{array}{c}
\sqrt{\dfrac{E^{(1)} + m_1}{2 \, E^{(1)}}} \; a_-(\vec k) \\[2.33ex]
-\sqrt{\dfrac{E^{(1)} - m_1}{2 \, E^{(1)}}} \; a_-(\vec k)
\end{array} \right) \,.
\end{align}
\end{subequations}
The negative-energy eigenstates of the tardyonic equations
in the helicity basis are given as
\begin{subequations}
\label{VV1}
\begin{align}
V^{(1)}_+(\vec k) =& \;
\frac{(m_1 - \cancel{k}) \, v_+(\vec k)}%
{\sqrt{(E^{(1)} - |\vec k|)^2 + m_1^2}} 
= \left( \begin{array}{c}
-\sqrt{\dfrac{E^{(1)} - m_1}{2 \, E^{(1)}}} \; a_+(\vec k) \\[2.33ex]
-\sqrt{\dfrac{E^{(1)} + m_1}{2 \, E^{(1)}}} \; a_+(\vec k)
\end{array} \right) \,,
\\[0.33ex]
V^{(1)}_-(\vec k) =& \;
\frac{(m_1 - \cancel{k}) \, v_-(\vec k)}%
{\sqrt{(E^{(1)} - |\vec k|)^2 + m_1^2}} 
= \left( \begin{array}{c}
-\sqrt{\dfrac{E^{(1)} - m_1}{2 \, E^{(1)}}} \; a_-(\vec k) \\[2.33ex]
\sqrt{\dfrac{E^{(1)} + m_1}{2 \, E^{(1)}}} \; a_-(\vec k)
\end{array} \right) \,.
\end{align}
\end{subequations}
These solutions are consistent with those given in 
Ref.~\cite{JaWi1959} and in Chap.~23 of Ref.~\cite{BeLiPi1982vol4}, 
and the normalizations are ($\sigma = \pm$)
\begin{equation}
U^{(1)\plus}_\sigma(\vec k) \, U^{(1)}_\sigma(\vec k) =
V^{(1)\plus}_\sigma(\vec k) \, V^{(1)}_\sigma(\vec k) = 1 \,.
\end{equation}
One can change the normalization according to 
\begin{align}
\calU^{(1)}_\sigma(\vec k) = 
\left( \frac{E^{(1)}}{m_1} \right)^{1/2} \, 
U^{(1)}_\sigma(\vec k) \,,
\qquad
\calV^{(1)}_\sigma(\vec k) =
\left( \frac{E^{(1)}}{m_1} \right)^{1/2} \, 
V^{(1)}_\sigma(\vec k) \,.
\end{align}
The Lorentz-invariant normalization is equal to one for the 
fundamental positive-energy bispinors and equal to minus one 
for the fundamental negative-energy bispinors,
\begin{equation}
\label{covariantUV1}
\overline \calU^{(1)}_\sigma(\vec k) \; 
\calU^{(1)}_\sigma(\vec k) = 1 \,,
\qquad
\overline \calV^{(1)}_\sigma(\vec k) \;
\calV^{(1)}_\sigma(\vec k) = -1 \,.
\end{equation}
A little algebra is sufficient to reproduce the following 
known sums over bispinors,
\color{\redcol}
\begin{subequations}
\label{spinorsum1}
\begin{align}
\fbox{$\displaystyle{\sum_\sigma \calU^{(1)}_\sigma(\vec k) \otimes
{\overline \calU}^{(1)}_\sigma(\vec k) =
\frac{\cancel{k} + m_1}{2 m_1} \,, }$}
\qquad
\fbox{$\displaystyle{
\sum_\sigma \calV^{(1)}_\sigma(\vec k) \otimes
{\overline \calV}^{(1)}_\sigma(\vec k) = 
\frac{\cancel{k} - m_1}{2 m_1} \,. }$}
\end{align}
\end{subequations}
\color{black}
In accordance with general wisdom about the
tardyonic case, these do not involve a helicity-dependent
prefactor. The sum rule~\eqref{spinorsum1} is of type I
[see~Eq.~\eqref{sumrule1}].

%
%
\subsection{Two Tardyonic Mass Terms}
\label{sec23}

Inspired by the discussion in Sec.~\ref{intro},
we consider an equation with two tardyonic mass terms,
which has already been indicated in Eq.~\eqref{eq1} and
reads
$\left( \ii \gamma^\mu \partial_\mu - m_1 - \ii \, \gamma^5 m_2 \right) 
\psi^{(t)}(x) =0$.
For the corresponding bispinors in the 
fundamental plane-wave solutions, this implies that 
\begin{subequations}
\begin{align}
\left( \cancel{k} - m_1 - \ii \gamma^5 \, m_2 \right) \, 
U^{(t)}_{\pm}(\vec k) =& \; 0 \,,
\\
\left( -\cancel{k} - m_1 - \ii \gamma^5 \, m_2 \right) \, 
V^{(t)}_{\pm}(\vec k) =& \; 0 \,.
\end{align}
\end{subequations}
The dispersion relation is $E^{(t)} = \sqrt{ \vec k^2 + m_1^2 + m_2^2}$.
The fundamental positive-energy bispinors read as follows,
\begin{subequations}
\label{UUt}
\begin{align}
U^{(t)}_+(\vec k) = & \;
\frac{(\cancel{k} + m_1 - \ii \gamma^5\,m_2) \, u_+(\vec k)}%
{\sqrt{(E^{(t)} - |\vec k|)^2 + m_1^2 + m_2^2}} 
= \left( \begin{array}{c}
\dfrac{m_1-\ii \, m_2+E^{(t)}-|\vec k|}%
{\sqrt{(E^{(t)} - |\vec k|)^2 + m_1^2 + m_2^2}} \; 
\dfrac{a_+(\vec k)}{\sqrt{2}} \\[0.33ex]
\dfrac{m_1-\ii \, m_2-E^{(t)}+|\vec k|}%
{\sqrt{(E^{(t)} - |\vec k|)^2 + m_1^2 + m_2^2}} \; 
\dfrac{a_+(\vec k)}{\sqrt{2}} \\
\end{array} \right) \,,
\\[0.33ex]
U^{(t)}_-(\vec k) = & \;
\frac{(\cancel{k} + m_1 - \ii \gamma^5\,m_2) \, u_-(\vec k)}%
{\sqrt{(E^{(t)} - |\vec k|)^2 + m_1^2 + m_2^2}} 
= \left( \begin{array}{c}
\dfrac{m_1+\ii \, m_2+E^{(t)}-|\vec k|}%
{\sqrt{(E^{(t)} - |\vec k|)^2 + m_1^2 + m_2^2}} \; 
\dfrac{a_-(\vec k)}{\sqrt{2}}  \\[0.33ex]
\dfrac{-m_1-\ii \, m_2+E^{(t)}-|\vec k|}%
{\sqrt{(E^{(t)} - |\vec k|)^2 + m_1^2 + m_2^2}} \; 
\dfrac{a_-(\vec k)}{\sqrt{2}} \\
\end{array} \right) \,.
\end{align}
\end{subequations}
The negative-energy eigenstates of the 
equation with two tardyonic mass terms are given as
\begin{subequations}
\label{VVt}
\begin{align}
V^{(t)}_+(\vec k) = & \;
\frac{(-\cancel{k} - \ii \gamma^5\,m_2 + m_1) \, v_+(\vec k)}%
{\sqrt{(E^{(t)} - |\vec k|)^2 + m_1^2 + m_2^2}} 
= \left( \begin{array}{c}
\dfrac{-m_1+\ii \, m_2+E^{(t)}-|\vec k|}%
{\sqrt{(E^{(t)} - |\vec k|)^2 + m_1^2 + m_2^2}} \; 
\dfrac{a_+(\vec k)}{\sqrt{2}} \\[0.33ex]
\dfrac{-m_1+\ii \, m_2-E^{(t)}+|\vec k|}%
{\sqrt{(E^{(t)} - |\vec k|)^2 + m_1^2 + m_2^2}} \; 
\dfrac{a_+(\vec k)}{\sqrt{2}} \\
\end{array} \right) \,,
\nonumber
\end{align}
for negative helicity (positive chirality in the massless limit)
and 
\begin{align}
V^{(t)}_-(\vec k) = & \;
\frac{(-\cancel{k} - \ii \gamma^5\,m_2 + m_1) \, v_-(\vec k)}%
{\sqrt{(E^{(t)} - |\vec k|)^2 + m_1^2 + m_2^2}} 
= \left( \begin{array}{c}
\dfrac{-m_1-\ii \, m_2+E^{(t)}-|\vec k|}%
{\sqrt{(E^{(t)} - |\vec k|)^2 + m_1^2 + m_2^2}} \; 
\dfrac{a_-(\vec k)}{\sqrt{2}} \\[0.33ex]
\dfrac{m_1+\ii \, m_2+E^{(t)}-|\vec k|}%
{\sqrt{(E^{(t)} - |\vec k|)^2 + m_1^2 + m_2^2}} \; 
\dfrac{a_-(\vec k)}{\sqrt{2}} \\
\end{array} \right) \,.
\nonumber
\end{align}
\end{subequations}
for positive helicity (negative chirality in the massless limit).
In the massless limit
(first $E^{(t)} \to |\vec k|$, then $m_2 \to 0$, and then $m_1 \to 0$),
we again reproduce the massless solutions,
$U^{(t)}_+(\vec k) \to u_+(\vec k)$, $U^{(t)}_-(\vec k) \to u_-(\vec k)$,
$V^{(t)}_+(\vec k) \to v_+(\vec k)$ and $V^{(t)}_-(\vec k) \to v_-(\vec k)$.
Of course, in the limit $m_1 \to 0$, one also has to 
expand the normalization denominators in powers of $m_1$.
For $m_2 =0$ the solutions~\eqref{UUt} and~\eqref{VVt} reduce to the solutions
of the ordinary Dirac equation in Eqs.~\eqref{UU1} and~\eqref{VV1}, and can be
expanded in $m_2$ to yield corrections
to the ordinary Dirac equation for small $m_2$, i.e $m_1 \gg m_2$.
The states are normalized with respect to the condition
\begin{equation}
U^{(t)\plus}_\sigma(\vec k) \, U^{(t)}_\sigma(\vec k) =
V^{(t)\plus}_\sigma(\vec k) \, V^{(t)}_\sigma(\vec k) = 1 \,.
\end{equation}
In the normalization
\begin{subequations}
\begin{align}
\calU^{(t)}_\sigma(\vec k) = & \;
\left( \frac{E^{(t)}}{m_1} \right)^{1/2} \, 
U^{(t)}_\sigma(\vec k) \,,
\qquad
\calV^{(t)}_\sigma(\vec k) = 
\left( \frac{E^{(t)}}{m_1} \right)^{1/2} \, 
V^{(t)}_\sigma(\vec k) \,,
\end{align}
\end{subequations}
the positive-energy solutions acquire a ``positive Lorentz-invariant norm'',
whereas the negative-energy solutions have ``negative Lorentz-invariant norm'',
\begin{equation}
\overline \calU^{(t)}_\sigma(\vec k) \;
\calU^{(t)}_\sigma(\vec k) = 1 \,,
\qquad
\qquad
\overline \calV^{(t)}_\sigma(\vec k) \;
\calV^{(t)}_\sigma(\vec k) = -1 \,.
\end{equation}
After some algebra, one can derive the following sums over bispinors,
\color{\redcol}
\begin{equation}
\label{spinorsumt}
\fbox{$\displaystyle{\sum_\sigma \calU^{(t)}_\sigma(\vec k) \otimes
{\overline \calU}^{(t)}_\sigma(\vec k) =
\frac{\cancel{k} + m_1 - \ii \, \gamma^5 \, m_2}{2 m_1}} \,,$}
\qquad
\fbox{$\displaystyle{\sum_\sigma \calV^{(t)}_\sigma(\vec k) \otimes
{\overline \calV}^{(t)}_\sigma(\vec k) =
\frac{\cancel{k} - m_1 + \ii \, \gamma^5 \, m_2}{2 m_1}} \,.$}
\end{equation}
\color{black}
These are easily identified as the positive- and negative-energy projectors.
The sum rules do not involve helicity-dependent 
prefactor and are of type I [see Eq.~\eqref{sumrule1}].

The solutions~\eqref{UUt} and~\eqref{VVt} approach the massless solutions 
if one replaces $m_2 \to 0$ first and then 
lets $m_1 \to 0$. They are thus useful for systems where the $m_1$ mass 
is greater than $m_2$. 
For $m_2 \gg m_1$, one would like to 
calculate solutions that approach the massless case
for the sequence $m_1 \to 0$, then $m_2 \to 0$.
These read as follows,
\begin{equation}
U'^{(t)}_\sigma(\vec k) = \ii \, \sigma \, U^{(t)}_\sigma(\vec k) \,,
\qquad
V'^{(t)}_\sigma(\vec k) = \ii \, \sigma \, V^{(t)}_\sigma(\vec k) \,.
\end{equation}
In comparison to the solutions~\eqref{UUt} and~\eqref{VVt},
they acquire a nontrivial phase factor.

%
%
%
\section{Generalized Dirac Equations: Tachyonic Mass Terms}
\label{sec3}

%
%
\subsection{Tachyonic Dirac Equation}
\label{sec31}

The tachyonic Dirac equation is given in Eq.~\eqref{eq5} and reads 
$\left( \ii \gamma^\mu \, \partial_\mu - \gamma^5 \, m \right) \, \psi(x) = 0$.  
The fundamental bispinors entering the equations 
fulfill the equations
\begin{equation}
\left( \cancel{k} - \gamma^5 \, m \right) \, U_{\pm}(\vec k) = 0 \,,
\qquad
\left( \cancel{k} + \gamma^5 \, m \right) \, V_{\pm}(\vec k) = 0 \,.
\end{equation}
Using $(\cancel{k} - \gamma^5 \, m) \, (\cancel{k} - \gamma^5 \, m) = k^2 + m^2$ and 
some algebra, the prefactors in the fundamental
bispinors (for positive energy) take a very simple form,
\begin{subequations}
\label{UU}
\begin{align}
U_+(\vec k) = & \;
\frac{(\gamma^5\,m- \cancel{k}) \, u_+(\vec k)}%
{\sqrt{(E - |\vec k|)^2 + m^2}} 
=
\left( \begin{array}{c}
\sqrt{\dfrac{|\vec k| + m}{2 \, |\vec k|}} \; a_+(\vec k) \\[2.33ex]
\sqrt{\dfrac{|\vec k| - m}{2 \, |\vec k|}} \; a_+(\vec k) \\
\end{array} \right) 
\,,
\\[0.77ex]
U_-(\vec k) = & \;
\frac{(\cancel{k} - \gamma^5\,m) \, u_-(\vec k)}%
{\sqrt{(E - |\vec k|)^2 + m^2}} 
=
\left( \begin{array}{c}
\sqrt{\dfrac{|\vec k| - m}{2 \, |\vec k|}} \; a_-(\vec k) \\[2.33ex]
-\sqrt{\dfrac{|\vec k| + m}{2 \, |\vec k|}} \; a_-(\vec k) \\
\end{array} \right)\,.
\end{align}
\end{subequations}
For negative energy, the solutions read as follows,
\begin{subequations}
\label{VV}
\begin{align}
V_+(\vec k) = & \;
\frac{(\gamma^5\,m+\cancel{k}) \, v_+(\vec k)}%
{\sqrt{(E - |\vec k|)^2 + m^2}} 
=
\left( \begin{array}{c}
-\sqrt{\dfrac{|\vec k| - m}{2 \, |\vec k|}} \; a_+(\vec k) \\[2.33ex]
-\sqrt{\dfrac{|\vec k| + m}{2 \, |\vec k|}} \; a_+(\vec k) \\
\end{array} \right)\,,
\\[0.33ex]
V_-(\vec k) = & \;
\frac{(-\cancel{k} - \gamma^5 \,m) \, v_-(\vec k)}%
{\sqrt{(E - |\vec k|)^2 + m^2}}
=
\left( \begin{array}{c}
-\sqrt{\dfrac{|\vec k| + m}{2 \, |\vec k|}} \; a_-(\vec k) \\[2.33ex]
\sqrt{\dfrac{|\vec k| - m}{2 \, |\vec k|}} \; a_-(\vec k) \\
\end{array} \right) \,.
\end{align}
\end{subequations}
The normalization condition is ($\sigma = \pm$)
\begin{equation}
U^\plus_\sigma(\vec k) \, U_\sigma(\vec k) =
V^\plus_\sigma(\vec k) \, V_\sigma(\vec k) = 1 \,.
\end{equation}
One can change to Lorentz-invariant normalization 
by a multiplication with $(|\vec k|/m)^{1/2}$, 
\begin{equation}
\label{changeit}
\calU_\sigma(\vec k) = \left( \frac{|\vec k|}{m} \right)^{1/2} \, U_\sigma(\vec k) \,, 
\;\;
\calV_\sigma(\vec k) = \left( \frac{|\vec k|}{m} \right)^{1/2} \, V_\sigma(\vec k) \,.
\end{equation}
The ``calligraphic'' spinors 
fulfill the following helicity-dependent normalizations,
\begin{equation}
\label{covariant}
\overline \calU_\sigma(\vec k) \; \calU_\sigma(\vec k) = \sigma \,,
\qquad
\quad
\overline \calV_\sigma(\vec k) \; \calV_\sigma(\vec k) = -\sigma \,,
\end{equation}
where we observe that $\sigma$ is a good quantum number
because the helicity operator commutes with the Hamiltonian~\eqref{HD5}.
The sum rule fulfilled by the fundamental plane-wave 
spinors is of type II [see Eq.~\eqref{sumrule2}].
For the positive-energy spinors, we have
\color{\bluecol}
\begin{subequations}
\label{spinorsum5}
\begin{equation}
\label{tensora}
\fbox{$\displaystyle{ 
\sum_\sigma (-\sigma) \; \calU_\sigma(\vec k) \otimes
\overline\calU_\sigma(\vec k) \,\gamma^5 =
\frac{\cancel{k} - \gamma^5 \, m}{2 m} \,, }$}
\end{equation}
where for the negative-energy spinors, the sum rule reads
\begin{equation}
\label{tensorb}
\fbox{$\displaystyle{ 
\sum_\sigma (-\sigma) \; \calV_\sigma(\vec k) \otimes
\overline\calV_\sigma(\vec k) \,\gamma^5 =
\frac{\cancel{k} + \gamma^5 \, m}{2 m} \,. }$}
\end{equation}
\end{subequations}
\color{black}
The expressions on the right-hand sides are the positive- and negative-energy 
projectors.

%
%
\subsection{Two Tachyonic Mass Terms}
\label{sec32}

We study the equation
$\left( \ii  \, \gamma^\mu \, \partial_\mu - \ii m_1 - \gamma^5 m_2 \right) \psi(x) = 0$,
as given in Eq.~\eqref{eqp}.
The fundamental spinors, which we denote as 
$U'_{\pm}(\vec k)$ and $V'_{\pm}(\vec k)$, fulfill the 
following equations,
\begin{subequations}
\begin{align}
\label{orig}
\left( \cancel{k} - \ii \, m_1 - \gamma^5 \, m_2 \right) \, U'_{\pm}(\vec k) = & \; 0 \,,
\\
\left( \cancel{k} + \ii \, m_1 + \gamma^5 \, m_2  \right) \, V'_{\pm}(\vec k) = & \; 0 \,.
\end{align}
\end{subequations}
The positive-energy solutions are obtained using the 
identity $(\cancel{k} - \ii  m_1 - \gamma^5 m_2)\, (\cancel{k} + \ii\, m_1 - \gamma^5 m_2)=
k^2 + m_1^2 + m_2^2$.
With $E' = \sqrt{ \vec k^2 - m_1^2 - m_2^2}$, they read as follows,
\begin{subequations}
\label{UUp}
\begin{align}
U'_+(\vec k) = &\;
\left( \begin{array}{c}
\dfrac{\ii \, m_1+ m_2-E'+|\vec k|}%
{\sqrt{(E' - |\vec k|)^2 + m_1^2 + m_2^2}} \;
\dfrac{a_+(\vec k)}{\sqrt{2}} \\[0.33ex]
\dfrac{\ii \, m_1+m_2+E'-|\vec k|}%
{\sqrt{(E' - |\vec k|)^2 + m_1^2 + m_2^2}} \;
\dfrac{a_+(\vec k)}{\sqrt{2}} \\
\end{array} \right) \,,
\\[0.33ex]
U'_-(\vec k) =&\;
\left( \begin{array}{c}
\dfrac{\ii m_1+ m_2+E'-|\vec k|}%
{\sqrt{(E' - |\vec k|)^2 + m_1^2 + m_2^2}} \;
\dfrac{a_-(\vec k)}{\sqrt{2}}  \\[0.33ex]
\dfrac{-\ii m_1-m_2+E'-|\vec k|}%
{\sqrt{(E' - |\vec k|)^2 + m_1^2 + m_2^2}} \;
\dfrac{a_-(\vec k)}{\sqrt{2}} \\
\end{array} \right) \,.
\end{align}
\end{subequations}
The negative-energy solutions for the tachyonic equation
with two mass terms, are given as
\begin{subequations}
\label{VVp}
\begin{align}
V'_+(\vec k) =& \;
\left( \begin{array}{c}
\dfrac{\ii m_1-m_2-E'+|\vec k|}%
{\sqrt{(E' - |\vec k|)^2 + m_1^2 + m_2^2}} \;
\dfrac{a_+(\vec k)}{\sqrt{2}} \\[0.33ex]
\dfrac{\ii m_1- m_2-E'-|\vec k|}%
{\sqrt{(E' - |\vec k|)^2 + m_1^2 + m_2^2}} \;
\dfrac{a_+(\vec k)}{\sqrt{2}} \\
\end{array} \right) \,,
\\[0.33ex]
V'_-(\vec k) = & \;
\left( \begin{array}{c}
\dfrac{-\ii m_1- m_2+E'-|\vec k|}%
{\sqrt{(E' - |\vec k|)^2 + m_1^2 + m_2^2}} \;
\dfrac{a_-(\vec k)}{\sqrt{2}} \\[0.33ex]
\dfrac{\ii m_1+ m_2+E'-|\vec k|}%
{\sqrt{(E' - |\vec k|)^2 + m_1^2 + m_2^2}} \;
\dfrac{a_-(\vec k)}{\sqrt{2}} \\
\end{array} \right) \,.
\end{align}
\end{subequations}
The normalization condition is 
$U'^\plus_\sigma(\vec k) \, U'_\sigma(\vec k) =
V'^\plus_\sigma(\vec k) \, V'_\sigma(\vec k) = 1$.
We use a definition of the
``calligraphic'' spinors analogous to Eq.~\eqref{changeit},
\begin{equation}
\calU'_\sigma(\vec k) = 
\left( \frac{|\vec k|}{m_2} \right)^{1/2} \, U'_\sigma(\vec k) \,,
\qquad
\calV'_\sigma(\vec k) = 
\left( \frac{|\vec k|}{m_2} \right)^{1/2} \, V'_\sigma(\vec k) \,.
\end{equation}
In analogy to Eq.~\eqref{spinorsum5}, a sum rule of type II [see Eq.~\eqref{sumrule2}]
is fulfilled by the fundamental plane-wave spinors,
\color{\bluecol}
\begin{equation}
\label{spinorsump}
\begin{split}
\fbox{$\displaystyle{
\sum_\sigma (-\sigma) \, \calU'_\sigma(\vec k) \otimes
{\overline \calU}'_\sigma(\vec k) \, \gamma^5 =
\frac{\cancel{k} + \ii \, m_1 - \gamma^5 \, m_2}{2 m_2}} \,,$}
\\[0.33ex]
\fbox{$\displaystyle{
\sum_\sigma (-\sigma) \, \calV'_\sigma(\vec k) \otimes
{\overline \calV}'_\sigma(\vec k) \, \gamma^5 =
\frac{\cancel{k} - \ii \, m_1 + \gamma^5 \, m_2}{2 m_2}} \,.$}
\end{split}
\end{equation}
\color{black}
We thus obtain the desired projectors onto 
positive- and negative-energy solutions
for the Dirac equation with two tachyonic mass
terms~\eqref{eqp}.
The generalized equation
$\left( \ii  \, \gamma^\mu \, \partial_\mu - \ii m_1 - \gamma^5 m_2 \right) \psi(x) = 0$
is fully compatible with the Clifford-algebra based approach 
recently described in Ref.~\cite{Pa2012}.

%
%
\section{Theorems for Generalized Dirac Fields}
\label{sec4}

%
%
\subsection{Spinor Sums and Time-Ordered Propagator}
\label{sec41}

Our central postulate regarding the quantized fermionic theory 
is that the time-ordered vacuum expectation value of the 
field operators should yield the time-ordered (Feynman) propagator,
which, in the momentum representation, is equal to the 
inverse of the Hamiltonian (upon multiplication with $\gamma^0$).
This postulate implies that under rather general 
assumptions regarding the mathematical form of the 
elementary field anticommutators,
sum rules have to be fulfilled by the tensor sums over the 
fundamental spinor solutions.
It is perhaps not surprising that these sum rules, are precisely 
of the form investigated in Secs.~\ref{sec2} and~\ref{sec3} of this paper.

For definiteness, we consider the solution of the tachyonic 
Dirac equation (Sec.~\ref{sec31}).
The generalization to other generalized Dirac equations is 
straightforward. We start from the field operator~\cite{JeWu2012epjc}
\begin{align}
\label{fieldop}
\psi(x) =& \;
\int \frac{\dd^3 k}{(2\pi)^3} \,
\frac{m}{E} \sum_{\sigma = \pm}
\left\{ b_\sigma(k) \, \calU_\sigma(\vec k) \,
\ee^{-\ii \, k \cdot x}  
+ d^\plus_\sigma(k) \, \calV_\sigma(\vec k) \,
\ee^{\ii \, k \cdot x} \right\} \,,
\nonumber\\[0.33ex]
k =& \; (E, \vec k)\,,
\qquad E = E_{\vec k} =  \sqrt{\vec k^2 - m^2 - \ii \, \epsilon} \,,
\end{align}
where $b_\sigma$ annihilates particles and $d^\plus_\sigma$ creates antiparticles.
The following anticommutators vanish,
\begin{subequations}
\begin{align}
\left\{ b_\sigma(k) , b_{\rho}(k') \right\} = & \;
\left\{ b^\plus_\sigma(k) , b^\plus_{\rho}(k') \right\} = 0 \,,
\\
\left\{ d_\sigma(k) , d_{\rho}(k') \right\} = & \;
\left\{ d^\plus_\sigma(k) , d^\plus_{\rho}(k') \right\} = 0 \,,
\end{align}
\end{subequations}
We assume the following general form for the nonvanishing anticommutators,
\begin{subequations}
\label{fg}
\begin{align}
\left\{ b_\sigma(k) , b^\plus_{\rho}(k') \right\} = & \;
f(\sigma, \vec k) \;
(2 \pi)^3 \, \frac{E}{m} \delta^3(\vec k - \vec k') \,
\delta_{\sigma\rho}\,,\\
\left\{ d_\sigma(k) , d^\plus_{\rho}(k') \right\} = & \;
g(\sigma, \vec k) \;
(2 \pi)^3 \, \frac{E}{m} \delta^3(\vec k - \vec k') \,
\delta_{\sigma\rho}\,,
\end{align}
\end{subequations}
with arbitrary $f$ and $g$ functions of the quantum numbers
$\sigma$ and $\vec k$. One might argue that since 
$f$ and $g$ must be dimensionless, they can depend only on the 
dimensionless arguments $\sigma$ and $\vec k/m$, but that is a 
detail of the discussion which we do not pursue any further.
Our only assumption concerns the fact that the field anticommutators
should be diagonal in the helicity and wave vector quantum 
numbers, leading to the corresponding Kronecker and Dirac-$\delta$'s.

We assume that the spin-matrix $\Gamma$ either constitutes a 
Lorentz scalar or a pseudo-scalar quantity,
which is a scalar under the proper orthochronous Lorentz group.
The time-ordered product of field operators reads as
\begin{align}
\label{time_ordered}
\left< 0 \left| T \, \psi(x) \, \overline{\psi}(y) \; \Gamma \right| 0 \right> 
=& \;
\int \frac{\dd^3 k}{(2 \pi)^3}
\frac{m}{E} \,
\bigl\{ \Theta(x^0 - y^0) \; 
\ee^{-\ii k \cdot (x-y)} \,
\sum_{\sigma = \pm} 
f(\sigma, \vec k) \; \calU_{\sigma}(\vec k) \otimes
\overline \calU_{\sigma}(\vec k) \, \Gamma 
\\
& \; 
- \Theta(y^0 - x^0) \; \ee^{\ii k \cdot (x-y)} \,
\sum_{\sigma = \pm} 
g(\sigma, \vec k) \; \calV_{\sigma}(\vec k) \otimes
\overline \calV_{\sigma}(\vec k) \Gamma \bigr\} \,.
\nonumber
\end{align}
This equation contains the same coefficient functions $f$ and $g$ 
that enter into Eq.~\eqref{fg}.
In order to proceed with the derivation of the propagator, 
we must postulate that the following sum rules hold,
\color{\bluecol}
\begin{align}
\label{postulates}
\sum_\sigma f(\sigma, \vec k) \; \calU_\sigma(\vec k) \otimes
\overline\calU_\sigma(\vec k) \,\Gamma =& \;
\frac{\cancel{k} - \gamma^5 \, m}{2 m} \,,
\qquad
\sum_\sigma g(\sigma, \vec k) \; \calV_\sigma(\vec k) \otimes
\overline\calV_\sigma(\vec k) \,\Gamma =
\frac{\cancel{k} + \gamma^5 \, m}{2 m} \,.
\end{align}
\color{black}
The sum rule~\eqref{postulates} is crucial
for the further steps in the derivation of the time-ordered propagator.
Introducing a suitable complex integral 
representation for the step function, one obtains
from Eq.~\eqref{time_ordered}, using Eq.~\eqref{postulates},
after a few steps which we do not discuss in further detail
[see also Eqs.~(3.169) and Eq.~(3.170) of Ref.~\cite{ItZu1980}]
\begin{align}
\label{deriv1}
\left< 0 \left| T \, \psi(x) \, \overline\psi(y) \Gamma \right| 0 \right> 
=& \;
\ii  \int \frac{\dd^3 k}{(2 \pi)^3} \frac{m}{E} \,
\int \frac{\dd k_0}{2 \pi} \,
\ee^{-\ii k_0 \cdot (x^0-y^0) + \ii \vec k \cdot (\vec x - \vec y)} \, 
\frac{\gamma^0 k_0 - \vec \gamma \cdot \vec k - \gamma^5 \, m}{2 m
(k_0 - E + \ii\epsilon)}
\nonumber\\[0.77ex]
& \; + \ii \int \frac{\dd^3 k}{(2 \pi)^3} \frac{m}{E} \,
\int \frac{\dd k_0}{2 \pi} \,
\ee^{-\ii k_0 \cdot (x^0-y^0) + \ii \vec k \cdot (\vec x - \vec y)} \, 
\frac{-\gamma^0 k_0 + \vec \gamma \cdot \vec k + \gamma^5 \, m}{2 m
(k_0 + E - \ii\epsilon)} \,.
\end{align}
The convention is that in any integrals $\int \dd^3 k$, 
the component $k_0$ is set equal to $E = \sqrt{\vec k^2 - m^2}$ in the integrand 
when it occurs in scalar products of the 
form $k \cdot (x-y)$ etc., but if the integral is over the full $\dd^4 k$, 
then the integration interval is the full $k_0 \in (-\infty, \infty)$.
With the convention which is adopted in many quantum field theoretical textbooks,
including Refs.~\cite{ItZu1980,PeSc1995},
we finally obtain the result
\begin{align}
\left< 0 \left| T \, \psi(x) \, \overline\psi(y) \Gamma \right| 0 \right> 
=& \; 
\ii  \int \frac{\dd^4 k}{(2 \pi)^4} 
\ee^{-\ii k \cdot (x-y)} \, 
\frac{\cancel{k} - \gamma^5 \, m }{k^2 + m^2 + \ii\epsilon} \,.
\end{align}
The tachyonic propagator $S_T$ is identified, under the integral sign, as
\begin{equation}
\label{ST2}
S_T(k) = \frac{1}{\cancel{k} - \gamma^5 \, (m + \ii\,\epsilon)} =
\frac{\cancel{k} - \gamma^5 \, m}{k^2 + m^2 + \ii \, \epsilon} \,.
\end{equation}
The sum rule~\eqref{spinorsum5} implies that the derivation 
is valid for the choice
\color{\bluecol}
\begin{equation}
\label{choice5}
f(\sigma, \vec k) = g(\sigma, \vec k) = -\sigma \,, 
\qquad 
\qquad 
\Gamma = \gamma^5 \,,
\end{equation}
\color{black}
in which case the relations given in Eq.~\eqref{postulates} are fulfilled.
Note that this observation does not imply that the choice~\eqref{choice5}
necessarily is the only one for which we are able to fulfill the 
postulates given in Eq.~\eqref{postulates}, but it is 
the only structurally simple choice that we have found.

For the egregiously simple choice~\eqref{choice5}, 
let us study the transition to the massless limit~\eqref{sumrule2} 
in some further detail. Indeed, in the limit $m \to 0$, 
the denominator of the spin sums in Eq.~\eqref{spinorsum5} vanishes, and 
a finite limit is obtained after multiplication with $2 m$,
\color{\bluecol}
\begin{subequations}
\label{tensorchiral}
\begin{align}
\label{tensorchirala}
\lim_{m \to 0} 
\sum_\sigma 2 m \, (-\sigma) \; \calU_\sigma(\vec k) \otimes
\overline \calU_\sigma(\vec k) \, \gamma^5 =& \; 
\cancel{k},
\\[0.33ex]
\label{tensorchiralb}
\lim_{m \to 0} 
\sum_\sigma 2 m \, (-\sigma) \; \calV_\sigma(\vec k) \otimes
\overline \calV_\sigma(\vec k) \, \gamma^5 =& \; 
\cancel{k}.
\end{align}
\end{subequations}
\color{black}
In order to compare the normalizations of the fundamental spinors 
in the massless limit, we calculate the following quantities,
\begin{subequations}
\label{as_implied}
\begin{align}
\label{aa}
\overline \calU_\sigma(\vec k) \, \gamma^5 \, \gamma^0 \,
\calU_\sigma(\vec k) =& \; -\sigma \, \frac{E}{m} \,,
\qquad
\overline u_\sigma(\vec k) \, \gamma^5 \, \gamma^0 \,
u_\sigma(\vec k) = -\sigma \,,
\\
\label{cc}
\overline \calV_\sigma(\vec k) \, \gamma^5 \, \gamma^0 \,
\calV_\sigma(\vec k) =& \; -\sigma \, \frac{E}{m} \,,
\qquad
\overline v_\sigma(\vec k) \, \gamma^5 \, \gamma^0 \,
v_\sigma(\vec k) = -\sigma \,,
\end{align}
\end{subequations}
where the fundamental spinors 
$u_\sigma$ and $v_\sigma$ of the massless equation have been 
given in Sec.~\ref{sec21}.
Observing that $E = | \vec k |$ in the 
massless limit, the identifications
$\sqrt{m} \, \calU_\sigma(\vec k) \to
\sqrt{| \vec k |} \, u_\sigma(\vec k)$ and
$\sqrt{m} \, \calV_\sigma(\vec k) \to
\sqrt{| \vec k |} \, v_\sigma(\vec k)$,
as implied by Eq.~\eqref{as_implied},
show that the identities~\eqref{tensorchiral} 
precisely reduce
to the sum rule~\eqref{sumrule2} in the massless limit.

%
%
\subsection{Generalized Field Anticommutators for Tardyonic and Tachyonic Fields}
\label{sec42}

For definiteness, we have considered the case of the tachyonic 
Dirac field in the above derivation.
The decisive observation is that the choice
\color{\bluecol}
\begin{equation}
\mbox{tachyonic choice:} 
\quad
f(\sigma, \vec k) = g(\sigma, \vec k) = -\sigma \,,
\quad
\Gamma = \gamma^5 \,,
\end{equation}
\color{black}
is consistent with both massive tachyonic fields discussed in Secs.~\ref{sec31}
and~\ref{sec32}, whereas the tardyonic choice
\color{\redcol}
\begin{equation}
\mbox{tardyonic choice:} 
\quad
f(\sigma, \vec k) = g(\sigma, \vec k) = 1 \,,
\quad
\Gamma = \mathbbm{1}_{4 \times 4} \,,
\end{equation}
\color{black}
yields the time-ordered propagator for both 
massive tardyonic fields discussed in Sec.~\ref{sec22} and~\ref{sec23}.
The nonvanishing anticommutators for tardyons take the 
simple form [cf.~Eq.~\eqref{fg}],
\color{\redcol}
\begin{align}
\label{tardyonic_anticom}
& \mbox{tardyonic anticommutators:} 
\nonumber\\
& \left\{ b_\sigma(k) , b^\plus_{\rho}(k') \right\} = 
(2 \pi)^3 \, \frac{E}{m} \delta^3(\vec k - \vec k') \,
\delta_{\sigma\rho}\,,
\qquad
\left\{ d_\sigma(k) , d^\plus_{\rho}(k') \right\} = 
(2 \pi)^3 \, \frac{E}{m} \delta^3(\vec k - \vec k') \,
\delta_{\sigma\rho}\,.
\end{align}
\color{black}
Again, compared with Eq.~\eqref{fg},
the nonvanishing anticommutators for tachyons take the simple form
\color{\bluecol}
\begin{align}
\label{tachyonic_anticom}
& \mbox{tachyonic anticommutators:} 
\nonumber\\
& \left\{ b_\sigma(k) , b^\plus_{\rho}(k') \right\} = 
(-\sigma) \,
(2 \pi)^3 \, \frac{E}{m} \delta^3(\vec k - \vec k') \,
\delta_{\sigma\rho}\,,
\qquad
\left\{ d_\sigma(k) , d^\plus_{\rho}(k') \right\} = 
(-\sigma) \,
(2 \pi)^3 \, \frac{E}{m} \delta^3(\vec k - \vec k') \,
\delta_{\sigma\rho}\,.
\end{align}
\color{black}
With these universal choices, the theory of the tardyonic and 
tachyonic spin-$1/2$ fields can be unified.
The time-ordered propagator is given as
\begin{align}
\label{ST}
\left< 0 \left| T \, \psi(x) \, \overline{\psi}(y) \Gamma \right| 0 \right> =& \;
\ii \, S(x-y) \,,
\qquad
S(x-y) = \int \frac{\dd^4 k}{(2 \pi)^4} \,
\ee^{-\ii k \cdot (x-y)} \, S(k) \,.
\end{align}
We use the sum rules for the tensor sums over fundamental spinors
given in Eq.~\eqref{spinorsum1} (for the tardyonic Dirac field),
Eq.~\eqref{spinorsumt} (for the tardyonic Dirac field with two 
mass terms), Eq.~\eqref{spinorsum5} (for the tachyonic Dirac field)
and in Eq.~\eqref{spinorsump} (for the Dirac field of imaginary mass).
Going through the exact same derivation as outlined above in between
Eqs.~\eqref{time_ordered} and~\eqref{ST2}, we obtain the following 
results for the time-ordered propagators of tardyonic and tachyonic fields.
For the tardyonic Dirac field (Sec.~\ref{sec22}), one has
\begin{equation}
\label{fp_m1_eps}
S^{(1)}(k) = \frac{1}{\cancel{k} - m_1 + \ii \epsilon}
= \frac{\cancel{k} + m_1}{k^2 - m_1^2 + \ii \, \epsilon} \,.
\end{equation}
For the tardyonic field with two mass terms (Sec.~\eqref{sec23}),
the Feynman propagator is easily found as
\begin{align}
\label{fp_m1m2_epseta}
S^{(t)}(k) =& \; \frac{1}{\cancel{k} - m_1 + \ii \epsilon -
\ii \, \gamma^5 \, (m_2 - \ii \, \eta)}
= \frac{\cancel{k} + m_1 -  \ii \, \gamma^5 \, m_2}%
{k^2 - m_1^2 - m_2^2 + \ii \, \epsilon} \,,
\end{align}
where $\epsilon$ and $\eta$ are infinitesimal imaginary parts.
Both tardyonic mass terms acquire an infinitesimal negative 
imaginary part, and the prefactor $m/E$ from Eq.~\eqref{fieldop} 
in the field operator needs to be replaced by $m_1/E^{(t)}$ where the 
tardyonic energy is 
$E^{(t)} = \sqrt{ \vec k^2 + m_1^2 + m_2^2}$.
For the tachyonic Dirac field (Sec.~\ref{sec31} and Ref.~\cite{JeWu2012epjc}),
one has the result given in Eq.~\eqref{ST2}.
Finally, for the Dirac field with two tachyonic mass terms, we have
\begin{equation}
\label{STp}
S'(k) = \frac{\cancel{k} + \ii \, m_1 - \gamma^5 m_2}%
{k^2 + m_1^2 + m_2^2 + \ii \, \epsilon} \,.
\end{equation}
In the latter case, the prefactor $m/E$ from Eq.~\eqref{fieldop}
in the field operator needs to be replaced by $m_1/E'$ where the
tachyonic energy is $E' = \sqrt{ \vec k^2 - m_1^2 - m_2^2}$.
For both tachyonic fields discussed here, the 
mass acquires an infinitesimal positive imaginary part,
as manifest in the results given in 
Eqs.~\eqref{ST2} and~\eqref{STp}.

%
%
\subsection{Tachyonic Gordon Identities}
\label{sec43}

It is useful to illustrate the derivation outlined 
above by exploring its connection to tachyonic Gordon 
identities. For definiteness, we again concentrate on the 
tachyonic Dirac equation discussed in Sec.~\ref{sec31}.
The matrix element of the vector current finds the 
following Gordon decomposition
for positive-energy spinors,
\begin{subequations}
\label{tachcalJdef}
\begin{align}
\overline \calU_{\pm}(\vec k') \, \gamma^\mu \,
\calU_{\pm}(\vec k') 
= \frac{1}{2m}
\overline \calU_{\pm}(\vec k') \gamma^5 \, \left[ \left( k'^\mu - k^\mu \right)  
+\ii \sigma^{\mu \nu} 
\left( k'_{\nu} + k_\nu \right) \right] \calU_{\pm}(\vec k) \,.
\nonumber
\end{align}
For negative-energy solutions, the identity reads as
\begin{align}
\overline \calV_{\pm}(\vec k') \gamma^\mu \calV_{\pm}(\vec k) 
=  -\frac{1}{2m}\overline \calV_{\pm}(\vec k') \, \gamma^5 \, 
\left[ \left( k'^\mu - k^\mu \right)  
+\ii \sigma^{\mu \nu} \left( k'_{\nu} + k_\nu \right) \right] \,
\calV_{\pm} (\vec k) \,.
\nonumber
\end{align}
\end{subequations}
For $k'=k$, one has 
\begin{subequations}
\label{tachcurrent}
\begin{align}
\label{J1}
\overline \calU_{\pm}(\vec k) \gamma^\mu \calU_{\pm}(\vec k) 
=& \; \frac{\ii}{m} \overline \calU_{\pm}(\vec k) \, \gamma^5 \,
\sigma^{\mu \nu} k_\nu \, \calU_{\pm}(\vec k) \,, \\
\overline \calV_{\pm}(\vec k) \gamma^\mu \calV_{\pm}(\vec k) 
=&\; -\frac{\ii}{m} \overline \calV_{\pm}(\vec k) \, \gamma^5 \,
\sigma^{\mu \nu} k_\nu \, \calV_{\pm}(\vec k) \,.
\end{align}
\end{subequations}
The matrix element of the axial current reads
\begin{subequations}
\label{tachaxialcurrentJ} 
\begin{align}
\overline \calU_{\pm}(\vec k') \, \gamma^5 \gamma^\mu \, \calU_{\pm}(\vec k) 
= - \frac{1}{2m} \overline \calU_{\pm}(\vec k') 
\left[ \left( k'^\mu + k^\mu \right)  
+ \ii \sigma^{\mu \nu} \left( k'_{\nu} - k_\nu \right)  \right] 
\calU_{\pm} (\vec k) \,,
\end{align}
whereas for negative-energy solutions
\begin{align}
\overline \calV_{\pm}(\vec k') \, \gamma^5 \gamma^\mu \, \calV_{\pm}(\vec k)
= \frac{1}{2m}\overline \calV_{\pm}(\vec k')
\left[ \left( k'^\mu + k^\mu \right)
+\ii \sigma^{\mu \nu} \left( k'_{\nu}
- k_\nu \right)  \right] \calV_{\pm}(\vec k) \,. 
\end{align}
\end{subequations}
For $k'=k$, this simplifies to
\begin{subequations}
\label{tachaxialcurrent}
\begin{align}
\label{J2}
\overline \calU_{\pm}(\vec k) \gamma^5 \gamma^\mu \calU_{\pm}(\vec k) 
=& \; -\frac{1}{m} \overline \calU_{\pm}(\vec k) \, k^\mu \, \calU_{\pm}(\vec k) \,, \\
\overline \calV_{\pm}(\vec k) \gamma^5 \gamma^\mu \calV_{\pm}(\vec k) 
=& \; \frac{1}{m} \overline \calV_{\pm}(\vec k) \, k^\mu \, \calV_{\pm}(\vec k) \,.
\end{align}
\end{subequations}
The results~\eqref{tachaxialcurrentJ}
and~\eqref{tachaxialcurrent} 
for the tachyonic {\em axial vector} current 
have a similar structure 
as the Gordon decomposition
for the tardyonic {\em vector} current obtained 
with the ordinary Dirac equation
[see Eq.~(2.54) of Ref.~\cite{ItZu1980}].
The role of the Dirac adjoint for the tardyonic 
case is taken over by the ``chiral adjoint''
$\overline \calU_{\pm}(\vec k) \, \gamma^5$ for the tachyonic particle.
Here, the designation ``chiral adjoint'' is inspired by the fact that 
$\overline \calU_{\pm}(\vec k) \, \gamma^5 \, \calU_{\pm}(\vec k) $ transforms as 
a pseudo-scalar under Lorentz transformations.

The structure of Eqs.~\eqref{tachcurrent} and~\eqref{tachaxialcurrent}
is somewhat peculiar with regard to parity. 
In Eq.~\eqref{J1}, an apparent vector current 
on the left-hand side appears to transform into 
an axial current on the right-hand side,
whereas in Eq.~\eqref{J2}, an apparent axial vector on
the left-hand side of the equation becomes what appears to be a
vector on the right-hand side. The reason lies in the more complicated
behavior of the tachyonic Dirac equation under parity as investigated
in Ref.~\cite{JeWu2012jpa}. Namely, the tachyonic Dirac 
equation~\eqref{eq5} contains a term which transforms
as a scalar under parity,
\begin{subequations}
\begin{equation}
\ii \, \gamma^\nu \, \partial_\mu 
\mathop{\longrightarrow}^{\mathcal{P}} 
\gamma^0 \left( \ii \, \gamma^0 \, \partial_0 + 
\ii \, \gamma^i \, (-\partial_i) \right) \, \gamma^0 =
\ii \, \gamma^\nu \, \partial_\mu 
\end{equation}
as well as a term which transforms as a pseudoscalar,
\begin{equation}
\label{trafo2}
\gamma^5 \, m
\mathop{\longrightarrow}^{\mathcal{P}} 
\gamma^0 \left( \gamma^5 \, m \right) \, \gamma^0 =
- \gamma^5 \, m \,.
\end{equation}
\end{subequations}
The mass term in the tachyonic Dirac equation 
is pseudoscalar and changes sign under parity.
Indeed, in Ref.~\cite{JeWu2012jpa},
the tachyonic Dirac equation has been shown to be 
separately $\calC \calP$ invariant, and $\calT$ invariant,
but not $\calP$ invariant, due to the change in the mass term.

In order to put this observation into perspective,
we recall that the entries of the electromagnetic field strength 
tensor are composed of axial vector components (magnetic $\vec B$ field), 
as well as vector components (electric $\vec E$ field).
The transformation properties of the electromagnetic 
field strength tensor under the proper orthochronous Lorentz 
group are nevertheless well-defined.

The transformation~\eqref{trafo2} can be interpreted
as a transformation $m \to -m$ under parity.
Thus, if we interpret the mass $m$ as a pseudoscalar quantity, 
then the right-hand sides of~\eqref{J1} and~\eqref{J2} 
transform as a vector and an axial vector, respectively.
It is the parity non-invariance of the mass term in the 
tachyonic Dirac equation which leads to the 
somewhat peculiar structure of Eqs.~\eqref{J1} and~\eqref{J2}.

%
%
\subsection{Helicity--Dependence and Gupta--Bleuler Condition}
\label{sec44}

The anticommutator relations for tardyons given 
in Eq.~\eqref{tardyonic_anticom} imply that both left-handed 
as well as right-handed helicity states,
for both particles as well as antiparticles, have 
positive norm.
However, the anticommutator relations for tachyons given 
in Eq.~\eqref{tachyonic_anticom} imply that 
right-handed particle as well as left-handed
antiparticle states acquire negative norm.
This is shown in Eqs.~(31) and~(32) of Ref.~\cite{JeWu2012epjc}.
Indeed, for one-particle states $| 1_{k, \sigma} \rangle = b^+_\sigma(k) | 0 \rangle$,
\color{\bluecol}
\begin{align}
\label{negnorm}
\left< 1_{k, \sigma} | 1_{k, \sigma} \right> =& \;
\left< 0 \left| b_\sigma(k) \,
b^+_\sigma(k) \right| 0 \right>
= \left< 0 \left| \left\{ b_\sigma(k),
b^+_\sigma(k) \right\} \right| 0 \right>
= (-\sigma) \, V \, \frac{E}{m} \,,
\end{align}
\color{black}
where $V = (2 \pi)^3 \, \delta^3(\vec 0)$ is the normalization
volume in coordinate space. The
Fock-space norm $\left< 1_{k, \sigma} | 1_{k, \sigma} \right>$
is negative for $\sigma = 1$.

For clarification, the corresponding Gupta--Bleuler condition should be
indicated explicitly.
In full analogy to the instructive discussion of the Gupta--Bleuler
mechanism for the photon field, as given in full clarity in Chap.~9b of 
Schweber's textbook~\cite{Sc1961},
we select the positive- and negative-frequency component of the 
($\sigma=1$)-component of the neutrino field operator,
\begin{subequations}
\begin{align}
\psi_{\sigma=1}^{(+)}(x) =& \;
\int \frac{\dd^3 k}{(2\pi)^3} \,
\frac{m}{E} \; b_{\sigma=1}(k) \; \calU_{\sigma=1}(\vec k) \;
\ee^{-\ii \, k \cdot x} \,,
\\[2ex]
\psi_{\sigma=1}^{(-)}(x) =& \;
\int \frac{\dd^3 k}{(2\pi)^3} \,
\frac{m}{E} \; d^\plus_{\sigma=1}(k) \; \calV_{\sigma=1}(\vec k) \;
\ee^{\ii \, k \cdot x} \,,
\end{align}
\end{subequations}
and postulate that it annihilates any physical Fock state $| \Psi \rangle$
of the tachyonic field,
\begin{equation}
\label{stronger}
\langle \Psi | \psi^{(-)}_{\sigma = 1}(x) = 
\psi_{\sigma=1}^{(+)}(x) \; |\Psi \rangle = 0 \,.
\end{equation}
These relations automatically imply that the 
Gupta--Bleuler condition also is realized in terms of the 
expectation value
\begin{equation}
\fbox{$\displaystyle{
\langle \Psi | \psi_{\sigma = 1}(x) | \Psi \rangle =
\langle \Psi | \psi^{(-)}_{\sigma = 1}(x) + \psi^{(+)}_{\sigma = 1}(x) | \Psi \rangle 
= 0 \,, }$}
\end{equation}
but the condition~\eqref{stronger} is stronger.
We recall that the Gupta--Bleuler condition on the photon field reads
$\langle \Psi_\gamma | \partial^\mu A_\mu | \Psi_\gamma \rangle = 0$,
where $|  \Psi_\gamma \rangle$ is a Fock state of the photon field.
As stressed in Schweber's book~\cite{Sc1961} on p.~246,
the condition 
$\langle \Psi_\gamma | \partial_\mu A^\mu | \Psi_\gamma \rangle = 0$
is not sufficient for the suppression of the 
longitudinal and scalar photons, but one must postulate that
$\partial^\mu A_\mu^{(+)} | \Psi_\gamma \rangle = 0$,
where $A_\mu^{(+)}$ is the positive-frequency component of the 
photon field operator. Because the tachyonic fermion, unlike the photon,
is not equal to its own antiparticle, we need two conditions,
given in Eq.~\eqref{stronger}.

A crucial question now concerns the possibility of reversing
the helicity-dependence, i.e., the question of whether or not
a different choice for the helicity-dependent factors 
in Eq.~\eqref{fg} exists that would imply negative 
norm for left-handed particles and right-handed antiparticles.
A related question is whether other tachyonic Dirac Hamiltonians exist
for which the choice $f(\sigma, \vec k) = g(\sigma, \vec k) = +\sigma$
instead of $(-\sigma)$ would fulfill our general postulate,
namely, the sum rule~\eqref{postulates}.
For reasons outlined in the following, we can ascertain
that this is not the case;
tachyonic spin-$1/2$ particles should always be left-handed.

The arguments supporting this conclusion are as follows.
First, if we assume that the tachyonic field fulfills a
sum rule of type II which for massless fields is
given in Eq.~\eqref{sumrule2}, then it is impossible
to replace $(-\sigma)$ by $(+\sigma)$ in the sum rule because
of the necessity to preserve a smooth massless limit.
The considerations in the text following Eq.~\eqref{sumrule2}
imply that if one were to replace $(-\sigma)$ by $(+\sigma)$ 
in the massless case, then one would violate the sum rule~\eqref{sumrule2}.
The second argument is obtained by explicit calculation.
We have checked that if one replaces 
$m \to -m$ in the tachyonic and imaginary-mass Dirac 
equations~\eqref{eq5} and~\eqref{eqp}, then the sum rules
fulfilled by the corresponding fundamental 
spinors still contain the characteristic factor $(-\sigma)$.
For the imaginary-mass Dirac equation, this result
is obtained in Ref.~\cite{Je2012imag}.
Intuitively, we can understand this result as follows:
The mass $m$ in the denominators of the right-hand
sides of Eqs.~\eqref{spinorsum5} and~\eqref{spinorsump}
is obtained as the modulus $\sqrt{m^2} = |m|$ and does 
not change if we replace $m \to -m$ in the superluminal Dirac equation.
The mass in the numerator of the right-hand
sides of Eqs.~\eqref{spinorsum5} and~\eqref{spinorsump}
changes sign, but this is consistent with the obvious change in the 
functional form of the  positive-energy and negative-energy projectors
as we change the sign of the mass term.
Again, this consideration supports the conclusion 
that we cannot invert the helicity-dependence by choosing a
different Hamiltonian; the factor $(-\sigma)$ persists.

The third argument comes from the tachyonic Gordon identities
discussed in Sec.~\ref{sec43}. We use the Gordon identity 
Eq.~\eqref{tachaxialcurrent} and the normalization~\eqref{covariant} 
to calculate the bispinor trace (with a $\gamma^0$ multiplied from the 
right) of the left-hand side of Eq.~\eqref{tensora},
\begin{subequations}
\begin{align}
\label{rel1}
\trc\left( \sum_\sigma (-\sigma) \; \calU_\sigma(\vec k) \otimes
\overline \calU_\sigma(\vec k) \gamma^5 \gamma^0 \right) 
=& \; \sum_\sigma (-\sigma) \; 
\overline \calU_\sigma(\vec k) \, \gamma^5 \, \gamma^{\mu = 0} \, \calU_\sigma(\vec k)
\nonumber\\
= & \; \sum_\sigma (-\sigma) \; 
\left( - \frac{k^0}{m} \right) \, 
\underbrace{ \overline \calU_\sigma(\vec k) \, \calU_\sigma(\vec k) }%
_{=\sigma} = 2 \, \frac{E}{m} \,.
\end{align}
The bispinor trace of the right-hand side of Eq.~\eqref{tensora} is 
\begin{equation}
\trc\left( \gamma^0 \, \frac{\cancel{k} - \gamma^5 \, m}{2 m} \right) =
\trc\left( \gamma^0 \, \frac{\cancel{k}}{2 m} \right) = 4 \, \frac{E}{2 m} 
= 2 \, \frac{E}{m} \,,
\end{equation}
\end{subequations}
which shows the consistency of the bispinor sum~\eqref{tensora}
with the Gordon decomposition~\eqref{tachaxialcurrent}.
If we were to replace $(-\sigma)$ by $(+\sigma)$, the two sides
of the relation~\eqref{rel1} would differ by a minus sign.
The bispinor trace of the left-hand side of Eq.~\eqref{tensorb} is
\begin{subequations}
\begin{align}
\label{final1}
\trc\left( \sum_\sigma (-\sigma) \; \calV_\sigma(\vec k) \otimes
\overline \calV_\sigma(\vec k) \gamma^5 \gamma^0 \right) 
=& \; \sum_\sigma (-\sigma) \; 
\overline \calV_\sigma(\vec k) \, \gamma^5 \, \gamma^{\mu = 0} \, \calV_\sigma(\vec k)
\nonumber\\
& = \; \sum_\sigma (-\sigma) \; 
\left( \frac{k^0}{m} \right) \, 
\underbrace{\overline \calV_\sigma(\vec k) \, \calV_\sigma(\vec k)}%
_{=-\sigma} = 2 \, \frac{E}{m} \,.
\end{align}
We have used the tachyonic Gordon decomposition for negative-energy 
states as given in Eq.~\eqref{tachaxialcurrent},
which differs from the positive-energy Gordon decomposition by a minus sign,
but an additional minus sign is obtained  from the 
Lorentz-invariant normalization of the negative-energy fundamental bispinors.
From the right-hand side of Eq.~\eqref{tensorb},
we have
\begin{equation}
\label{final2}
\trc\left( \gamma^0 \, \frac{\cancel{k} + \gamma^5 \, m}{2 m} \right) =
\trc\left( \gamma^0 \, \frac{\cancel{k}}{2 m} \right) = 2 \, \frac{E}{m} \,,
\end{equation}
\end{subequations}
which again is fully consistent, but only because we have
a $(-\sigma)$ in the sum rule~\eqref{spinorsum5}, which combines with the 
$(-\sigma)$ from the Lorentz-invariant normalization of the 
$\overline \calV_\sigma(\vec k) \, \calV_\sigma(\vec k)$,
to give $2E/m$ as a final result in Eqs.~\eqref{final1} and~\eqref{final2}.

%
%
\section{Physical Interpretation: From Tachyonic Neutrinos to Cosmology}
\label{sec5}

%
%
\subsection{Arguments for and against Tachyonic Neutrinos}
\label{sec51}

In the absence of conclusive experimental evidence, 
the hypothesis of tachyonic neutrinos has been 
controversially discussed in the literature.
Three main arguments~\cite{HuSt1990} have been brought forward
against tachyonic neutrinos.
{\bf (1.)}~They would require us to give up the notion of
a Lorentz-invariant vacuum state, and even the vacuum 
would become unstable in the presence of tachyonic fields.
{\bf (2.)}~Given the tachyonic dispersion relation 
$E = \sqrt{\vec k^2 - m^2}$,
the role of states with $|\vec k| < m$ needs to be 
clarified.
{\bf (3.)}~The physical (probability!?) interpretation 
of the conserved Noether current 
of the free tachyonic Dirac equation has been called 
into question~\cite{HuSt1990}, and it has been argued 
that no consistent interpretation can be given
because certain zero components of the conserved current
were conjectured to vanish for all tachyonic momentum 
eigenstates~\cite{HuSt1990}.

A possible
answer for question {\bf (1.)}~has been proposed in Ref.~\cite{JeWu2012epjc}.
Summarizing the argument, it has been concluded in Ref.~\cite{JeWu2012epjc} that
one can solve the problem in two ways.  (i)~One can Lorentz transform the
vacuum state, and Lorentz transform all fundamental creation and annihilation
operators of the fermion field (some of these will change from annihilators to
creators upon transformation, due to the space-like nature of the tachyons).
(ii)~One keeps a Lorentz-invariant vacuum state, and only transform the
space-time {\em arguments} $k^\mu$ and $x^\mu$ 
of the field operators, keeping all creation
operators as creators and annihilation operators as annihilators.  The
amplitudes, cross sections, etc.~obtained using approach (ii)~then depend on
scalar product of four-vectors which are equal to the result obtained by first
calculating the process in the original Lorentz frame, and then, performing the
Lorentz transformation into the moving frame. 

The conjecture regarding an expansion about a ``false'' vacuum in the presence of
tachyons can be traced to the fact that most of the tachyonic theories
discussed so far in the literature are 
scalar~\cite{BiDeSu1962,Fe1967,ArSu1968,DhSu1968,BiSu1969,Fe1977}.
Indeed, scalar tachyons have a
problem with instability, because of the structure of the mass term in relation
to the field Hamiltonians, which changes sign $m^2 \to -m^2$ in a tachyonic
theory, suggesting that the field energy can be lowered by creating tachyons.
The problem does not occur in tachyonic spin-$1/2$ theories because a linear,
not quadratic, mass term enters the field Lagrangian and Hamiltonian.
Provided one reinterprets the spin-$1/2$ antiparticle solutions 
in the usual way (negative energy for propagation into the 
past becomes positive energy for propagation into the future),
it then becomes immediately clear that the vacuum energy cannot be 
lowered upon spin-$1/2$ tachyon anti-tachyon pair production.

An solution to problem~{\bf (2.)} has also been proposed in 
Ref.~\cite{JeWu2012epjc}. Namely, the energies with
$E = \pm \sqrt{\vec k^2 - m^2 - \ii \, \epsilon}$ 
(for $|\vec k| < m$) find a natural 
interpretation in terms of complex resonance 
and antiresonance energies, which describe unstable
states which decay in time. Particle resonances are damped 
for propagation into the future, antiparticle antiresonances
are damped for propagation into the past, as they should be~\cite{JeWu2012epjc}.
The occurrence of momentum eigenstates with real energies,
and resonances with complex resonance energies, 
is a well-known phenomenon all across physics (e.g., in atomic 
physics, the ground-state energy of the helium atom is strictly 
real, whereas auto-ionizing resonances in the three-particle 
system have a manifestly complex resonance energy).

Another ``myth'' which should be refuted concerns a conceivable
``runaway reaction'' where a moving tachyon releases an arbitrarily large amount
of energy, as it loses energy and accelerates, given its classical
energy-velocity relation $E = m/\sqrt{v^2 - 1}$.
According to this relation, a tachyon indeed accelerates as
its energy is lowered and becomes commensurate with the 
invariant mass square. (At high energy, a tachyon approaches
the light cone, though.) An infinitely fast tachyon takes the 
role of a tardyon at rest~\cite{BiDeSu1962}; the explicit eigenstates
have been indicated in Ref.~\cite{JeWu2012epjc}.
However, energy conservation holds, and one cannot mix the notion
of energy increase by acceleration, which only holds 
for tardyonic particles, in order to ``convert'' a tachyon 
losing energy by acceleration into a tardyon that gains 
energy in the same process. The ``runaway reaction''
is impossible; only a finite amount of energy is released
as the tachyon accelerates and the 
energy goes from $E = m/\sqrt{v^2 - 1}$ to zero.
Energy conservation holds for tachyons,
even if they accelerate, somewhat counterintuitively,
when losing energy.

An answer to question {\bf (3.)}
has not yet been provided so far in the literature
to the best of our knowledge.
Here, we aim to provide a possible physical interpretation
for the conserved current and scalar product,
and also show where certain arguments presented
originally by Hughes and Stephenson in their 
research article~\cite{HuSt1990} entitled ``against tachyonic neutrinos''
become inconsistent.
The Lagrangian density of the 
tachyonic Dirac particle reads, in first quantization~\cite{ChHaKo1985},
\begin{equation}
\label{Lchiral}
\calL(x) = 
\overline\psi(x) \, \gamma^5 \, 
\left( \ii \gamma^\mu \partial_\mu - \gamma^5 \, m \right)
\psi(x) \,.
\end{equation}
This is equivalent, up to partial integration, to the 
symmetric form~\cite{JeWu2012jpa}
\begin{equation}
\calL = 
\frac{\ii}{2} \left( \overline\psi \, \gamma^5 \, \gamma^\mu (\partial_\mu \psi) -
(\partial_\mu \overline\psi) \, \gamma^5 \, \gamma^\mu \psi   \right) - 
m \, \overline\psi \, \psi \,,
\end{equation}
where we suppress the space-time argument $x = (t, \vec r)$.
The non-symmetric form~\eqref{Lchiral} clearly exhibits the 
presence of the ``chiral adjoint'' $\overline\psi(x) \, \gamma^5$
in the Lagrangian. The conserved current is
\begin{equation}
\calJ^\mu(x) =
\overline\psi(x) \, \gamma^5 \, \gamma^\mu \, \psi(x) \,,
\qquad
\partial_\mu \calJ^\mu(x) = 0 \,.
\end{equation}
The zero component 
\begin{equation}
\calJ^0(x) =
\overline\psi(x) \, \gamma^5 \, \gamma^0 \, \psi(x) 
\end{equation}
assumes the following values in plane-wave eigenstates,
according to Eqs.~\eqref{aa} and~\eqref{cc},
\begin{subequations}
\label{77}
\begin{align}
\label{77a}
\overline \calU_\sigma(\vec k) \, \gamma^5 \, \gamma^0 \,
\calU_\sigma(\vec k) =& \; -\sigma \, \frac{E}{m} \,,
\\
\label{77b}
\overline \calV_\sigma(\vec k) \, \gamma^5 \, \gamma^0 \,
\calV_\sigma(\vec k) =& \; -\sigma \, \frac{E}{m} \,.
\end{align}
\end{subequations}
We conclude that $\calJ^0$ is positive for left-handed particle states
($\sigma  = -1$) and right-handed antiparticle states
(likewise, $\sigma  = -1$; the helicity is equal to $-\sigma$
for antiparticles).
For $E=0$, the ``axial norm'' of the momentum eigenstates
given in Eq.~\eqref{77} vanishes, but this is a 
very special case. Namely, for tachyons, $E=0$ implies 
$|\vec k| = m = m \, v/\sqrt{v^2-1}$ for the momentum and 
thus corresponds to an ``infinitely fast'' tachyon ($v \to \infty$),
which remains ``infinitely fast'' upon Lorentz transformation.
It would thus be wrong to conclude, as done in 
Ref.~\cite{HuSt1990} in the text following Eq.~(54) of Ref.~\cite{HuSt1990},
that all tachyonic states have zero axial norm,
because this would correspond to a forbidden generalization 
of a result which holds asymptotically, for $v \to \infty$,
to all finite values of the tachyonic velocity $v$.
Our Eq.~\eqref{77} gives the explicit result 
for any finite value of the energy $E$.

The spatial integral of the zero component of the conserved current,
\begin{align}
\int \dd^3 r \, \calJ^0(x) =
\int \dd^3 r \; \overline \psi(x) \, \gamma^5 \, \gamma^\mu \, \psi(x)
= -\int \dd^3 r \; \psi^\plus(x) \, \gamma^5 \, \psi(x) \,,
\end{align}
is precisely equal (up to a sign) to the 
scalar product $\langle \psi_1(t), \psi_2(t) \rangle \equiv
\int \dd^3 r \, \psi^\plus_1(t, \vec r) \, \gamma^5 \, \psi_2(t, \vec r)$
which is conserved under the time-evolution 
by the $\gamma^5$ Hermitian (pseudo-Hermitian) Hamiltonian
$H_5$.  This scalar product is not positive definite,
as already noticed in the work of Pauli~\cite{Pa1943},
and precisely corresponds to the scalar product introduced 
by Pauli in Eq.~(3) of Ref.~\cite{Pa1943}, where in the 
notation of Ref.~\cite{Pa1943} we have $\eta = \gamma^5$.
Similar observations have been made in Eqs.~(34) and~(52) of Ref.~\cite{HuSt1990}.
According to Refs.~\cite{BeBrReRe2004,BeBrChWa2005CP},
one could otherwise define a so-called $\calC$ operator,
which is not equal to the charge conjugation operator,
and ``remedies'' the problem of negative norm 
attained by some states under the $\gamma^5$ norm,
leading to a redefined, positive-definite scalar 
product. However, the negative norm 
finds a rather natural interpretation, and a redefinition
of the scalar product $\langle \psi_1(t), \psi_2(t) \rangle \equiv
\int \dd^3 r \, \psi^\plus_1(t, \vec r) \, \gamma^5 \, \psi_2(t, \vec r)$
therefore is not required.

For the ordinary Dirac equation, the 
conserved current is $J^\mu(x) = \overline\psi_e(x) \, \gamma^\mu \,\psi_e(x)$.
Its timelike component is $J^0(x) = \psi_e^\plus(x) \,\psi_e(x)$,
where the subscript $e$ reminds us of the electron.
The latter can be interpreted as a positive-definite 
probability density which is  conserved under the time evolution
generated by the ordinary Hermitian Dirac Hamiltonian 
$H^{(1)}$. 

The tachyonic Dirac current $\calJ^\mu$ is obtained 
from the ordinary Dirac $J^\mu$ by the replacement
$\overline \psi_e(x) \to \overline \psi(x) \, \gamma^5$.
We have seen that the scalar product for the tachyonic Dirac Hamiltonian
is equal to an integral over the timelike component of the 
conserved Noether current of the Dirac equation 
and is not positive-definite.
In order to put this observation into
perspective, it is instructive to recall that 
for the Klein--Gordon equation 
$(\partial_\mu \partial^\mu + m^2) \phi(x) = 0$, the 
zero component of the conserved current 
$j^\mu(x) = \frac{\ii}{2 m} \left( \phi^*(x) \partial_\mu \phi(x) - 
\phi(x) \partial^\mu \phi^*(x) \right)$ is not positive-definite, either.
Therefore, the zero component of the 
Klein-Gordon current cannot be 
interpreted as a probability density but must 
be interpreted as a charge density,
which is positive for particles and negative for antiparticles. 

This interpretation is not available 
for the zero component of the Noether current of the 
tachyonic Dirac equation, because the equation is 
not charge conjugation invariant and is primarily proposed to 
describe neutrinos~\cite{JeWu2012jpa}.
However, we can come closer to a physical interpretation of the 
timelike component of the Noether current of the
tachyonic equation if we compare the interaction Lagrangian
$\calL_{\rm QED}$
of quantum electrodynamics to the weak interaction 
$\calL_W$ of a 
neutrino and a ``heavy photon'', i.e., a $Z^0$ boson,
\begin{align}
\calL_{\rm QED} =& \; -e \, \overline \psi_e \, \gamma^\mu \, \psi_e \, A_\mu 
\qquad
\leftrightarrow
\qquad
\calL_W = -\frac{e}{2 \, \sin\theta_W \, \cos\theta_W}
\overline \psi \, \left( \gamma^\mu \, \frac{1 - \gamma^5}{2} \right) \psi \, Z_\mu \,,
\end{align}
where $\theta_W$ is the Weinberg angle.
The axial vector part $\calL^A_W$ of $\calL_W$ is
\begin{equation}
\calL^A_W 
= -\frac{e}{2 \, \sin(2 \theta_W)} \overline \psi \; \gamma^5 \gamma^\mu \; \psi \; Z_\mu,
\end{equation}
where the ordering of the $\gamma$ matrices is important.
For the interaction with the time-like component of the vector potential,
we have the expressions
\begin{align}
\calL^0_{\rm QED} =&\; -e \, \overline \psi_e \, \gamma^0 \, \psi_e \, A_0
\qquad
\leftrightarrow
\qquad
\calL^{A,0}_W 
= -\frac{e}{2 \, \sin(2\theta_W)} \overline \psi \; \gamma^5 \gamma^0 \; \psi \; Z_0.
\end{align}
For quantum electrodynamics (QED), 
we interpret $\overline \psi_e \, \gamma^0 \, \psi_e = 
\psi^\plus_e \, \psi_e$ as the probability density, and this suggests an
interpretation of the expression
$ \overline \psi \; \gamma^5 \gamma^0 \; \psi $ as an
``axial probability density'' or ``axial interaction density'' of the
neutrino field with the timelike component of the $Z^0$ boson.
According to Eqs.~\eqref{77a} and~\eqref{77b},
the ``axial interaction density'' is positive for the physically allowed 
states (left-handed particle and right-handed antiparticle states),
and negative for the physically forbidden states
(right-handed particle and left-handed antiparticle states).
This consideration is independent of the suppression
mechanism for the states of ``wrong'' helicity 
which, in second quantization, 
due to negative norm [see the discussion following 
Eq.~\eqref{negnorm}].

In general, the physical interpretation of tachyonic
theories has been discussed by Feinberg~\cite{Fe1967,Fe1977} and
Sudarshan {\em et al.}~\cite{BiDeSu1962,ArSu1968,DhSu1968,BiSu1969}.
The reinterpretation principle is a cornerstone of the
theory. The only physically sensible quantities in a quantum
theory are transition amplitudes. If the time-ordering of
superluminal events changes upon Lorentz transformation, then
one reinterprets the amplitude as connecting two space-time
events whose coordinates are transformed according to the
Lorentz transformation, so that the only physically sensible
quantity (transition amplitude) simply connects two events one of which happens
before the other~\cite{BiSu1969}. 
This may seem counter-intuitive at first, but
it is perhaps a little less counter-intuitive if we
take into account that the accepted formulation
of quantum field theory is based on the Feynman propagator
and on the reinterpretation principle for the
``advanced'' part of the Feynman Green function,
which propagates anti-particle solutions
into the past. In order to avoid problems with regard to
causality, the canonical quantum field theory of subluminal
particles has to be supplemented by a reinterpretation principle,
just like the tachyonic theory.
The physical interpretation of the Noether current 
is not affected by this consideration.

Finally, let us point out that even without invoking 
reinterpretation, superluminal propagation can be compatible 
with causality if we postulate that the tachyonic mass $m$
is so small that the superluminality is within the limits 
set forth by the uncertainty relation.
With an energy $E = m/\sqrt{\beta^2 - 1} \approx m/\sqrt{2 \delta}$ 
with $v = 1 + \delta$ and $\delta \ll 1$,
we have $\Delta E \, \Delta t \approx m/\sqrt{2 \delta} \, \Delta t\leq 1$
(in units with $\hbar = 1$).
Thus, for an infinitesimal tachyonic mass parameter
which does not exceed $m \leq \sqrt{2 \delta}/\Delta t$, 
causality is preserved within the limits set by the uncertainty principle.
Quantum limitations, the role of unstable modes and quantum tunneling
in superluminal propagation have been discussed in
Refs.~\cite{AhReSt1998,EnNi1992,NiSt2008,Ni2009}.

%
%
\subsection{Gupta--Bleuler Condition and Seesaw Mechanism}
\label{sec52}

The commonly accepted mechanism for the 
suppression of right-handed neutrino and left-handed 
antineutrino states is the seesaw mechanism~\cite{BiGi2012}.
After integrating the heavy degrees of freedom 
(the sterile neutrino), it contains
a nonrenormalizable dimension-five operator
and has a hierarchy problem: Namely, the neutrino 
masses are inversely proportional to the 
grand unification (GUT) scale $\Lambda$ and sensitively depend
on fine-tuning of $\Lambda$. Let us consider 
Eq.~(18) of Ref.~\cite{BiGi2012} and {\em hypothetically}
consider a formulation that would result if 
the physically observable neutrino were right-handed. 
Then, we could reformulate Eq.~(18) of Ref.~\cite{BiGi2012} as
\begin{equation} 
\calL_I^{\rm eff} = -\frac{1}{\Lambda} \,
\sum_{r,r'} \left[ {\bar R}_{r'R} \tilde H \right] 
Y_{r' r} \, \left[ {\tilde H}^T \, \left( R_{rR}\right)^c \right] + 
\mbox{h.c.},
\end{equation} 
where
\begin{equation}
R_{rR} = \left( \begin{array}{c} \nu_{rR} \\ r_R \end{array} \right) \,,
\qquad
H =  \left( \begin{array}{c} H^{(+)} \\ H^{(0)} \end{array} \right) \,.
\end{equation}
With the expectation value of the Higgs field,
\begin{equation}
\tilde H_0 = \frac{1}{\sqrt{2}} \, \left( \begin{array}{c} v \\ 0 \end{array} \right) 
\,,
\end{equation}
instead of the left-handed  Majorana neutrino masses,
the right-handed ones would be small,
\begin{align}
\calL^{\rm M} = -\frac12 \, \sum_{r,r'} {\bar \nu}_{r'R} M^R_{r r'} \, 
\left( \nu_{rR} \right)^c + \mbox{h.c.} \,,
\qquad
M^R_{r r'} = \frac{v^2}{\Lambda} \, Y_{r'r} \,.
\end{align}
The seesaw mechanism is not unique in suppressing 
a definite helicity of the neutrino. It is unique 
provided one formulates it in terms of the 
left-handed fermion fields, but it could be formulated
with inverted helicities if the helicity of the
observed neutrinos at low energy were different.
In the latter case, the right-handed neutrino
mass would be small, instead of the left-handed one.
On the other hand, the seesaw mechanism has the 
distinct advantage that it is not necessary to assume
a superluminal character of the neutrino.

The mechanism discussed here in the text following Eq.~\eqref{negnorm}
is definite in making a prediction regarding the 
suppression of {\rm right-}handed neutrino states,
as explained by three independent arguments in Sec.~\ref{sec44}.
However, we have to assume a superluminal neutrino.
While an interacting superluminal field theory is 
problematic, it is perhaps not as problematic 
as previously thought 
(see Ref.~\cite{Tr2012} and Sec.~4 of~Ref.~\cite{JeWu2012epjc}).

Final clarification can only come from 
experiment. The seesaw mechanism is compatible 
with a Majorana neutrino. The tachyonic Dirac 
equation implies that the neutrino cannot be 
equal to its antiparticle, because it does not allow
charge-conjugation invariant solutions.
It is only $\calC \calP$, but not $\calC$ invariant.
Experimental evidence for neutrinoless double beta decay
is disputed~\cite{KKEtAl2004},
and direct measurements of the neutrino mass square 
currently exclude neither positive nor 
negative values~\cite{RoEtAl1991,AsEtAl1994,StDe1995,AsEtAl1996,%
WeEtAl1999,LoEtAl1999,BeEtAl2008}.
The generally accepted seesaw mechanism implies that 
neutrino masses are generated by a nonrenormalizable 
interaction with a concomitant hierarchy problem,
and the mechanism in itself could be reformulated 
with opposite helicities. It is compatible with a 
Majorana neutrino. By contrast, 
a tachyonic neutrino is ``automatically'' left-handed
and not equal to its own antiparticle.
It is described by a $\gamma^5$ Hermitian Hamiltonian 
and is plagued with the conceptual difficulties 
associated with (ever so slightly) superluminal 
propagation. This means that it is experimentally possible
to test the models.

%
%
\subsection{Tachyonic Neutrinos as a Candidate for Dark Energy}
\label{sec53}

The formulation of a gravitational interaction of a spin-$1/2$ particle 
is nontrivial in the quantized formalism. 
Brill and Wheeler~\cite{BrWh1957} performed the 
pioneering steps in this direction. 
In order to formulate the gravitational coupling of a
Dirac particle, one has to formulate generalized 
Dirac matrices $\overline \gamma^\mu$, 
which fulfill anticommutation relations compatible
with the local metric $\overline g^{\mu\nu}(x)$ of curved 
space-time. Based on the Christoffel symbols
${\Gamma^\rho}_{\mu\nu} = {\Gamma^\rho}_{\mu\nu}(x)$, 
one formulates the Christoffel affine connection 
matrices $\Gamma_\mu$ in spinor space and 
calculates the covariant derivative $\nabla_\mu$ as 
follows~\cite{BrWh1957,Bo1975prd,SoMuGr1977,ShCa1991,ShVa1992},
\begin{align}
& \{ \overline\gamma^\mu(x), \overline\gamma^\mu(x) \} 
= 2\,\overline g^{\mu\nu}(x) \,,
\qquad
\qquad
{\Gamma^\rho}_{\mu\nu} =
\frac12 \, \overline g^{\rho\sigma}
\left( 
\frac{\partial \overline g_{\nu\sigma}}{\partial x^\mu} +
\frac{\partial \overline g_{\mu\sigma}}{\partial x^\nu} -
\frac{\partial \overline g_{\mu\nu}}{\partial x^\sigma} 
\right) \,,
\\[2ex]
& \nabla_\nu \overline \gamma_\mu =
\frac{\partial \overline \gamma_\mu}{\partial x^\nu} -
\Gamma^\rho_{\mu\nu} \overline\gamma_\rho +
\overline \gamma_\mu \, \Gamma_\nu -
\Gamma_\nu \, \overline\gamma_\mu = 0 \,,
\qquad
\qquad
\Gamma_\mu = - \frac14 \, \overline \gamma^\nu \,
\left( \frac{\partial \overline \gamma_\mu}{\partial x^\nu } - 
\overline \gamma_\sigma \, \Gamma^\sigma_{\nu\mu} \right) \,.
\end{align}
The above formula for $\Gamma_\mu$ is valid in the 
case of a diagonal metric $\overline g^{\mu\nu}$ 
such as the Schwarzschild metric considered in Ref.~\cite{SoMuGr1977}.
For a general space-time metric, and with ``West--Coast'' conventions 
for the local vierbein $g^{\mu\nu} = {\rm diag}(1,-1,-1,-1)$
(``East--Coast'' conventions were used in Ref.~\cite{BrWh1957}),
the result reads as 
\begin{equation} 
\label{solution}
\Gamma_k = - \frac{\ii}{4} \, \overline g_{\mu\alpha} \, 
\left( \frac{\partial {b_\nu}^\beta}{\partial x^k} \, 
{a^\alpha}_\beta - 
{\Gamma^\alpha}_{\nu k} \right) \, 
 \overline\sigma^{\mu\nu} \,,
\end{equation}
where $\overline\sigma^{\mu\nu} = 
\frac{\ii}{2} \, [ \overline \gamma^\mu, \overline \gamma^\nu ]$
is the spin matrix.
The ${a^\alpha}_\beta$ and ${b_\nu}^\beta$ coefficients
transform the Dirac matrices to the local vierbein,
\begin{equation} 
\overline \gamma_\rho = {b_\rho}^\alpha \, \gamma_\alpha \,,
\qquad
\gamma_\rho = {a^\alpha}_\rho \, \overline\gamma_\alpha \,,
\qquad
\overline \gamma^\alpha = {a^\alpha}_\rho \, \gamma^\rho\,,
\qquad
\gamma^\alpha = {b_\rho}^\alpha \, \overline\gamma^\rho \,.
\end{equation} 
With the covariant derivative $\nabla_\mu = \partial_\mu - \Gamma_\mu$, 
the gravitationally coupled Dirac equation reads 
as~\cite{BrWh1957,Bo1975prd,SoMuGr1977}
\begin{equation}
\label{gravtarddirac}
\left( \ii \, \overline\gamma^\mu \, \nabla_\mu - m \right) \psi(x) = 0 \,.
\end{equation}
In the case of a tachyonic Dirac particle, it has 
to be reformulated as follows,
\begin{equation}
\label{gravtachdirac}
\fbox{$\displaystyle{
\left( \ii \, \overline\gamma^\mu \, \nabla_\mu 
- \overline\gamma^5(x) \, m \right) \psi(x) = 0 \,.}$}
\end{equation}
The space-time coordinate-dependent matrix $\overline\gamma^5(x)$
can be defined as 
\begin{equation}
\label{gamma5}
\overline\gamma^5(x) 
= \frac{\ii}{4!} \, 
\frac{\overline\varepsilon_{\alpha\beta\gamma\delta}}{\sqrt{-\overline g}}  \, 
\overline \gamma^\alpha \, \overline \gamma^\beta \,
\overline \gamma^\gamma \, \overline \gamma^\delta 
= \frac{\ii}{4!} \; \epsilon_{\alpha\beta\gamma\delta} \;
\overline \gamma^\alpha \, \overline \gamma^\beta \,
\overline \gamma^\gamma \, \overline \gamma^\delta 
= \ii \; \overline \gamma^0 \, \overline \gamma^1 \,
\overline \gamma^2 \, \overline \gamma^3 \,,
\end{equation}
where $\overline g = {\rm det} \, \overline g^{\mu\nu}$ is the determinant of the 
metric, and $\overline\varepsilon_{\alpha\beta\gamma\delta} = 
\sqrt{-\overline g} \; \varepsilon_{\alpha\beta\gamma\delta}$
is the local $\epsilon$ tensor, while $\varepsilon_{\alpha\beta\gamma\delta}$
is the totaly antisymmetric Levi--Civit\`{a} tensor.
The last identity in Eq.~\eqref{gamma5} is valid for 
a diagonal metric~$\overline g^{\mu\nu}$.

Let us first discuss Eq.~\eqref{gravtarddirac} very briefly.
A projection onto the upper and lower radial components $f(r)$ and 
$g(r)$ in a gravitational field can be found in 
Eqs.~(19) and (20)~of Ref.~\cite{SoMuGr1977}.
For vanishing electrostatic potential $V\to 0$,
Eq.~(20) of Ref.~\cite{SoMuGr1977} is invariant 
under the replacement $f(r) \leftrightarrow g(r)$, 
and $E \leftrightarrow -E$.
So, if $E$ is an eigenvalue of the gravitationally coupled 
Dirac equation, so is $-E$.
Invoking reinterpretation and replacing $-E \to E$
for antiparticles,
we find that the spectrum of the gravitationally coupled
Dirac Hamiltonian is the same for particles and antiparticles.
Therefore, the formalism makes the unique prediction that 
tardyonic antiparticles, like tardyonic
particles, are attracted by a gravitational field. 
In passing, we note that the often 
cited motivation for the investigation of trapped antihydrogen and 
its interaction with the gravitational field therefore 
is faced with a unique 
theoretical prediction: Antiparticles are attracted by 
gravitation as much as particles are.

It has been confirmed within the last 
two decades~\cite{EfSuMa1990,HoPr1991,Pe1998,RiEtAl1998} that the 
Universe expands more rapidly on large distance scales 
than compatible with the matter density in the 
Universe. Coupling to a scalar field (``quintessence'') 
is usually invoked in order to explain 
the expansion of the expansion rate of the Universe~\cite{St2003}.
As the scalar quintessence field ``rolls down its potential'',
it accelerates the expansion rate of the Universe.
Because of a self-attracting property 
(the quintessence field energy can be lowered
by increasing the local density of the quintessence field),
quintessence has positive energy density 
but negative pressure. This property is necessary in order to 
constitute a candidate for dark energy~\cite{St2003}.
In the standard model of cosmology, 
dark energy accounts for about $73$\% of the 
total mass-energy of the Universe.
The quintessence field thereby acts like a time-dependent
cosmological constant.

We can thus conclude that most of the energy in the Universe actually is not
gravitationally attractive, i.e., that gravity can repel.  In the following, we
shall present qualitative arguments which suggest that tachyonic neutrinos may
play a role in the expansion of the Universe, and may contribute to ``dark
energy''. Conceivable connections of tachyonic physics and dark energy have been
explored in the literature, employing either scalar fields~\cite{BaJaPa2003} or
Lorentz-violating mechanisms~\cite{CiEvLiZh2011}.  In order to provide an
alternative explanation for dark energy, it is necessary to invoke a mechanism
that leads to a repulsive gravitational force on intergalactic distance scales
in the Universe.  It is not fully surprising that tachyonic neutrinos may
provide for such an alternative mechanism.  Namely, both on the classical
level~\cite{ReMi1974}, as well as on the level of quantum theory
[Eq.~\eqref{gravtachdirac}], tachyonic particles are repulsed by gravitational
fields.

On the classical level, this is seen as follows.
We start from the familiar equation of motion of a particle 
with mass $m$ in curved space-time [see Eq.~(86') of Ref.~\cite{ReMi1974}],
which is a geodesic,
\begin{equation}
\frac{\dd^2 x^\mu}{\dd^2 s} + 
{\Gamma^\mu}_{\rho\sigma} \, \frac{\dd x^\rho}{\dd s} \,
\frac{\dd x^\sigma}{\dd s} = 0 \,.
\end{equation}
With a suitably redefined proper time $\dd s'$, the 
zero geodesic for a tachyon reads as [see Eq.~(86') of Ref.~\cite{ReMi1974}]
\begin{equation}
\frac{\dd^2 x'^\mu}{\dd^2 s'} + 
{\Gamma'^\mu}_{\rho\sigma} \, \frac{\dd x'^\rho}{\dd s'} \,
\frac{\dd x'^\sigma}{\dd s'} = 0 \,.
\end{equation}
The force exerted on a tardyon reads,
according to Eq.~(79a) of Ref.~\cite{ReMi1974},
\begin{equation}
F^\mu = m \, \frac{\dd^2 x^\mu}{\dd s^2} 
= - m \, {\Gamma^\mu}_{\rho\sigma} \, 
\frac{\dd x^\rho}{\dd s} \, \frac{\dd x^\sigma}{\dd s} \,.
\end{equation}
However, there is a sign change for a tachyon 
[see Eq.~(79b) of Ref.~\cite{ReMi1974}],
\begin{equation}
F'^\mu = - m \, \frac{\dd^2 x'^\mu}{\dd s'^2}
= m \, {\Gamma'^\mu}_{\rho\sigma} \, 
\frac{\dd x'^\rho}{\dd s} \, \frac{\dd x'^\sigma}{\dd s} \,,
\end{equation}
corresponding to the change in the energy-velocity relation
$E = m/\sqrt{1 - v^2} \to E = m/\sqrt{v^2-1}$
for tardyon versus tachyon,
which leads to gravitational repulsion for tachyons.
Another intuitive way to understand gravitational tachyonic repulsion
is to observe that the quantity
$E = \sqrt{\vec k^2 + m^2}$ gets smaller when the 
position-dependent mass $m= m(x)$ gets smaller,
whereas the tachyonic energy $E = \sqrt{\vec k^2 - m^2}$ gets larger when the 
position-dependent mass $m= m(x)$ gets smaller.

As neutrinos are ubiquitous within the cosmos, one now needs to evaluate their
conceivable contribution to the repulsive force on intergalactic distance
scales. A detailed discussion is beyond the scope of the current article,
but we can formulate some initial considerations and
order-of-magnitude estimates.  One can
identify, a priori, two sources of neutrinos which may become important,
(i)~thermalized neutrinos which decouple in the early
Universe~\cite{HaMa1995,LoDoHeTu1999,TrMe2005,Ha2005}, and
(ii)~non-thermalized, high-energy neutrinos which are continuously generated
through various cosmic processes, such as nuclear fusion in stars like the sun,
supernovae, and high-energy cosmic background.

It is generally assumed that today's cosmic background (CMB) radiation is
accompanied by a neutrino background, which at some point (low energy)
decoupled from the other particles. 
The average energy of the background neutrinos is generally assumed to be of
the same order-of-magnitude as compared to the cosmic microwave background
(CMB), and the currently accepted value for the temperature of the neutrino
background is $(4/11)^{1/3} = 0.714$ times that of the electromagnetic
background radiation, i.e., about $1.9 \, {\rm K}$. 
The Universe, according to this hypothesis, would be filled with a sea of
nonrelativistic background neutrinos.  This picture changes when we assume that the
neutrino is tachyonic.  In the earliest stages of the  Universe, tachyonic
neutrinos are of course highly energetic, with $E = \sqrt{\vec k^2 -
m^2} \approx |\vec k | \gg 0$. As they lose energy, they become faster,
because $E = m/\sqrt{v^2-1} \to 0$ (in the classical
theory). The energy $E = \sqrt{\vec k^2 - m^2} $ vanishes as $|\vec k| \to m$,
within quantum theory. 
For $|\vec k| < m$, according to~\cite{JeWu2012epjc}, the spectrum
of the tachyonic Hamiltonian contains unstable resonances and anti-resonances,
with a purely imaginary resonance energy.
Because the real part of the resonance energy 
$E = -\ii \sqrt{m^2 - \vec k^2}$ vanishes,
the unstable states form a ``background'' of fluctuating 
quantum states in the Universe.
Particle resonances ($E = -\ii \sqrt{m^2 - \vec k^2}$) 
and antiparticle antiresonances ($E = +\ii \sqrt{m^2 - \vec k^2}$) are damped
for propagation into the future and past, respectively
(see Ref.~\cite{JeWu2012epjc}). The latter case amounts to propagation into the 
future if one invokes reinterpretation
for the antiparticle solutions~\cite{JeWu2012epjc}.

A very course-grained estimate regarding the role of the 
fluctuating resonance and antiresonance states can be given 
as follows. We start from the grand canonical emsemble
\begin{equation}
\Omega = - 2 \frac{k_B \, T}{2 \pi^2} \, V \,
\int_0^\infty \dd k \, k^2 \, 
\ln\left( 1 + \exp\left( \frac{\mu - E(k)}{k_B T} \right) \right) \,,
\end{equation}
where $k_B$ is the Boltzmann constant, $V$ is the normalization
volume, $\mu$ is the chemical potential,  and
$E(k)$ is the energy of a tachyon as a function of $k = |\vec k|$.
The multiplicity factor $2$ takes into account the particle and 
antiparticle resonances and is supplemented here in 
comparison to the derivation presented in Ref.~\cite{Tr2012tach}.
Furthermore, we have assumed an isotropic 
energy dependence $\int \dd^3 k \to 4 \pi \int \dd k \, k^2$. 
The pressure is found as
\begin{equation}
p = - \frac{\Omega}{V} = \frac{k_B \, T}{\pi^2} \, 
\int_0^\infty \dd k \, k^2 \, 
\ln\left( 1 + \exp\left( \frac{\mu - E(k)}{k_B T} \right) \right) \,.
\end{equation}
For $T \to 0$, the contribution of resonances and antiresonances 
with $|\vec k| < m$ to the pressure and energy density 
is easily evaluated according to Eqs.~(27) and (28) of Ref.~\cite{Tr2012tach}.
For the pressure, we have
\begin{equation}
\label{p0}
p_0 
= \frac{\ii}{3 \pi^2} \,\int_0^m \dd k \, k^3 \, \frac{\dd {\rm Im} \, E(k)}{\dd k}
= \frac{\ii}{3 \pi^2} \,\int_0^m \dd k \, \frac{k^4}{\sqrt{m^2 - k^2}}
= \frac{\ii \, m^4}{16 \, \pi} \,,
\end{equation}
whereas the energy density is
\begin{equation}
\label{rho0}
\rho_0 
= \frac{\ii}{\pi^2} \,\int_0^m \dd k \, k^2 \, {\rm Im} E(k)
= -\frac{\ii}{\pi^2} \,\int_0^m \dd k \, k^2 \, \sqrt{m^2 - k^2} 
= -\frac{\ii \, m^4}{16 \, \pi} = -p_0 \,.
\end{equation}
The equation of state fulfilled by the zero-temperature limit of the 
tachyonic Dirac sea is $w_0 = p_0/\rho_0 = -1$.
This is the required equation of state for a ``vacuum''
energy density that describes a 
nonvanishing cosmological constant~\cite{cosmo}.
If $p_0 + \rho_0 = 0$, then there is no net energy gain 
upon pulling on a ``piston'' which contains the Universe 
(an illustrative discussion can be found in Ref.~\cite{astro}).

Both results in Eqs.~\eqref{p0} and~\eqref{rho0} are
imaginary. This raises the pertinent question of how to 
incorporate the imaginary pressure and energy 
density of the neutrino field into the evolution
equations of the Universe. In giving rise to 
unstable resonance states, the tachyonic Dirac field,
if it exists, would be different from any other known fundamental quantum
fields~\cite{ItZu1980}. An exhaustive mathematical description would require an 
extension of scattering theory to complex energy 
and momentum exchanges, applied to low-energy collisions
with neutrinos, and describe the continuous decay and repopulation
of the unstable tachyonic states in the Universe.
A detailed discussion is beyond the scope of the current article.
However, in a first approximation, we can argue that
an established technique for the mathematical treatment of complex
resonances entails complex scaling
(for reviews, see Refs.~\cite{Mo1998,CaEtAl2007}),
and we thus explore a complex scaling ansatz to the 
solution of the evolution equations in the following.

The imaginary results in 
Eqs.~\eqref{p0} and~\eqref{rho0} describe the quantum fluctuations
due to the unstable states. 
We thus consider Eqs.~(8)--(11) of Ref.~\cite{cosmo},
which relate the accelerated expansion rate $\ddot a/|a|$ and the 
Hubble constant $H = \dot a/|a|$ (the dot denotes differentiation
with respect to time) to the energy densities of various 
cosmologically relevant quantities,
\begin{subequations}
\label{hubble}
\begin{align}
\label{hubble_a}
H^2 = \left( \frac{\dot a}{|a|} \right)^2 = & \;
\frac{8 \pi G}{3} \, \left( \rho_\Lambda + \rho_M + \rho_k \right) 
\approx \frac{8 \pi G}{3} \, \left( \rho_\Lambda + \rho_M \right) \,,
\\[2ex]
\label{hubble_b}
\frac{\ddot a}{|a|} = & \;
\frac{8 \pi G}{3} \, \rho_\Lambda -
\frac{4 \pi G}{3} \, \left( \rho_M + 3 p_M +\rho_k + 3 p_k \right) 
\approx 
\frac{4 \pi G}{3} \, \left( 2 \rho_\Lambda - \rho_M \right) \,.
\end{align}
\end{subequations}
Here, $\rho$ is the matter density in the Universe, 
$\rho_k = -3 k/(8 \pi G a^2)$ is the energy
density associated with the curvature of the Universe
($k = +1,0,-1$), while $\rho_\Lambda$ is the 
energy density corresponding to the cosmological constant.
The mass density is $\rho_M$ and $p_M \approx 0$.
We aim to solve Eq.~\eqref{hubble} using a complex 
variable ansatz $a = |a| \, \exp(\ii \theta)$ and $t = t \, \exp(\ii \varphi)$.
Using Eq.~\eqref{hubble}, with a value 
$H = 71 \, {\rm km} \; {\rm s}^{-1} \; {\rm Mpc}^{-1}$
for the Hubble constant, one easily reproduces
the accepted value for the critical mass density of
$\rho_{\rm crit} \approx 9.34 \times 10^{-33} \, {\rm kg} \; {\rm cm}^{-3}$
(in SI units) which corresponds to a 
critical energy density of
$\rho_{\rm crit} \approx 8.39 \times 10^{-9} \, {\rm J} \; {\rm m}^{-3}$
(we have set $c=1$). In natural units,
the result converts to $\rho_{\rm crit} \approx 9.99 \times 10^{-45} \; \GeV^4$.

We courageously set $k=0$ in Eq.~\eqref{hubble} (flat Universe) and 
match the modulus of the (complex) accelerated expansion $\ddot a/|a|$
against the currently accepted values of 
$\Omega_M = \rho/\rho_{\rm crit} \approx 0.27$
(see Ref.~\cite{DoVo1999}) and
$\Omega_\Lambda = \rho_\Lambda/\rho_{\rm crit} \approx 0.73$.
According to Eq.~\eqref{hubble_b},
the matching is performed by courageously assuming that the 
bulk of the cosmological constant is given by the 
complex energy density~\eqref{rho0}, and equating the 
complex modulus as follows,
\begin{equation}
\rho'_\Lambda \approx \rho_0 =
-\frac{\ii \, m^4}{16 \, \pi}  \,,
\qquad
\left| 2 \rho'_\Lambda - \rho_M \right| = 
\sqrt{ 4 \, | \rho'_\Lambda|^2 + \rho_M^2}  
\; \mathop{=}^{\mbox{!}} \;
2 \rho_\Lambda - \rho_M \approx 1.19 \, \rho_{\rm crit} \,.
\end{equation}
The solution is $\rho'_\Lambda = -\ii \, 0.579 \; \rho_{\rm crit}$,
and so our estimate for the neutrino mass reads as
\begin{equation}
\label{tachmass}
(5.79 \times 10^{-45} \, \GeV^4) \sim | \rho_0 | = \frac{m^4}{16 \pi}
\qquad
\Rightarrow
\qquad
\fbox{$\displaystyle{ m \sim 0.0232 \, \eV\,.}$}
\end{equation}
If this treatment is valid, then the mass $m$ necessarily has to 
be the mass of the ``heaviest'' neutrino mass eigenstate,
i.e., the one with the largest modulus of the tachyonic mass $m$.
Heavier eigenstates would lead to a larger cosmological constant,
because $\rho_0$ is proportional to the fourth power of the neutrino 
mass. In terms of a conversion to flavor eigenstates,
the heaviest mass eigenstates might well be the one
closest to the electron neutrino and
electron antineutrino lepton flavor eigenstates~\cite{Bi2006}.

We have interpreted the quantities
$p_0$ and $\rho_0$ given in Eqs.~\eqref{p0} and~\eqref{rho0}
as contributions to a cosmological constant which 
describes the evolution of the Universe in complex 
space and time directions $a = |a| \, \exp(\ii \theta)$ and 
$t = t \, \exp(\ii \varphi)$.
We should supplement the solutions for the 
complex rotation angles for space and time, which 
read as $\theta = 38.1^\circ$ and $\varphi = 70.6^\circ$,
within our complex scaling approach.
The evolution of the energy density with the 
aging of the Universe is given by 
[see Eq.~(12) of Ref.~\cite{cosmo}]
\begin{equation}
\rho_0 = \rho_i(0) \, a^{-3 \, (1 + w_0)} = {\mathrm{const.}} \,,
\qquad
w_0 = -1 \,.
\end{equation}
We therefore obtain a time-independent
energy density $\rho_0$, which is consistent with the fundamental 
character of the fluctuating unstable resonances of the neutrino 
field. We reemphasize that the mass $m \sim 0.0232 \, \eV$ is 
tachyonic and enters the 
gravitationally coupled tachyonic Dirac equation~\eqref{gravtachdirac}.
The given mass value is not currently excluded by any terrestrial experiment,
and it is consistent with the observed time spread of the arrival
times of neutrinos from the supernova SN1987A (see Ref.~\cite{ArRo1987}).
The hypothesis that very light elementary spin-$1/2$ particles
could be tachyonic has been pursued elsewhere~\cite{AKDy2003}.

%
%
\section{Conclusions}
\label{sec6}

In the current investigation,
we present the fundamental solutions of 
generalized Dirac equations in the helicity basis,
in a systematic and unified manner.
Of particular importance are the Dirac equation with 
two tardyonic mass terms $m_1 + \ii \, \gamma^5 \, m_2$ and
two tachyonic mass terms $\ii m_1 + \gamma^5 \, m_2$.
Let us summarize the main results.
We have discussed the ordinary (tardyonic) Dirac equation 
in Sec.~\ref{sec22}, a tardyonic Dirac equation (with two 
mass terms) in Sec.~\ref{sec23}, and two tachyonic Dirac
equations in Secs.~\ref{sec31} and~\ref{sec32}.
We give the fundamental eigenspinors that enter the
plane-wave solutions of all of these equations 
[see Eqs.~\eqref{UU1},~\eqref{VV1},~\eqref{UUt},~\eqref{VVt},
\eqref{UU},~\eqref{VV},~\eqref{UUp} and~\eqref{VVp}].
The eigenspinors are obtained using projector techniques
as outlined in Chap.~2 of Ref.~\cite{ItZu1980}.
For the ``normal'' Dirac equation
[see Eqs.~\eqref{UU1} and~\eqref{VV1}], our results are consistent 
with Ref.~\cite{JaWi1959} and Chap.~23 of Ref.~\cite{BeLiPi1982vol4}.
For the generalized Dirac equations, the solutions have
not yet appeared in the literature in the compact form given in the current 
article, to the best of our knowledge.
For the ordinary (tardyonic) Dirac equation, for the 
tachyonic Dirac equation, and for the imaginary-mass Dirac
equation, the prefactors are brought into compact
analytic form [see Eqs.~\eqref{UU1},~\eqref{VV1},
\eqref{UU},~\eqref{VV},~\eqref{UUp} and~\eqref{VVp}].

Finally, in Secs.~\ref{sec41} and~\ref{sec42}, we find that 
the tardyonic and tachyonic theories can be unified on the 
basis of the structurally simple anticommutator 
relations given in Eqs.~\eqref{tardyonic_anticom} 
and~\eqref{tachyonic_anticom}, which are independent of
the magnitude of the mass terms.
As outlined in Sec.~\ref{sec41}, the coefficient functions
$f = f(\sigma, \vec k)$ and $g = g(\sigma, \vec k)$ in the 
postulated form of the anticommutators~\eqref{fg} enter the 
tensor sums over the fundamental spinors in Eq.~\eqref{postulates}.
For the egregiously simple choices indicated in Eqs.~\eqref{tardyonic_anticom}
and~\eqref{tachyonic_anticom}, which are
consistent with a smooth massless limit,
the tensor sums over the fundamental eigenspinors 
yield the positive-energy and negative-energy projectors,
for both tardyonic as well as tachyonic eigenspinors.
Consistency with the massless limit requires the 
presence of the factor $(-\sigma)$ in the tensor sums 
over the eigenspinors for the tachyonic equations.
This fact is verified, on the basis of tachyonic Gordon 
identities and related considerations, in Secs.~\ref{sec43} and~\ref{sec44}.
The presence of the factor $(-\sigma)$ in the fundamental
tachyonic field anticommutators in
Eq.~\eqref{tachyonic_anticom}
implies the suppression of right-handed particle and left-handed
antiparticle states, due to negative norm,
as shown in Eq.~\eqref{negnorm}.

Finally, in Sec.~\ref{sec5}, we observe that since tachyons 
are repelled by gravity, it might be worthwhile to investigate 
their conceivable role in the mechanism(s) responsible for the 
accelerated expansion of the Universe (``dark energy'').
Furthermore, the tachyonic resonances and anti-resonance
energies  might play a role in the sum over states that
enters the thermodynamic potentials of a free tachyonic 
fermionic gas in the low-temperature limit.
If we consider the states with imaginary energy 
to be unstable, fluctuating states, then it is
intuitively obvious that they might contribute to
a fluctuating energy density and pressure on large 
distance scales in the Universe. As described in a somewhat 
approach in Sec.~\ref{sec53}, an order-of-magnitude 
calculation based on a complex scaling 
transformation of the time evolution of the Universe 
leads to a tachyonic neutrino mass which is not 
excluded at present by any terrestrial experiment.
We believe that it might be worthwhile to 
explore the physical consequences of the 
``ugly duckling'' (tachyonic neutrino) somewhat further.
It allows us to retain, among other things, 
lepton number conservation as a symmetry of nature.

%
%
\section*{Acknowledgments}

The authors acknowledge helpful conversations with C. Hirata, 
R.~J.~Hill, W. Rodejohann, P. J. Mohr, and I.~N\'{a}ndori.
This work was supported by the NSF (grant PHY-1068547) and by the
National Institute of Standards and Technology
(precision measurement grant).\\

\begin{center}
\begin{minipage}{14cm}
\begin{center}
\includegraphics[width=1.0\linewidth]{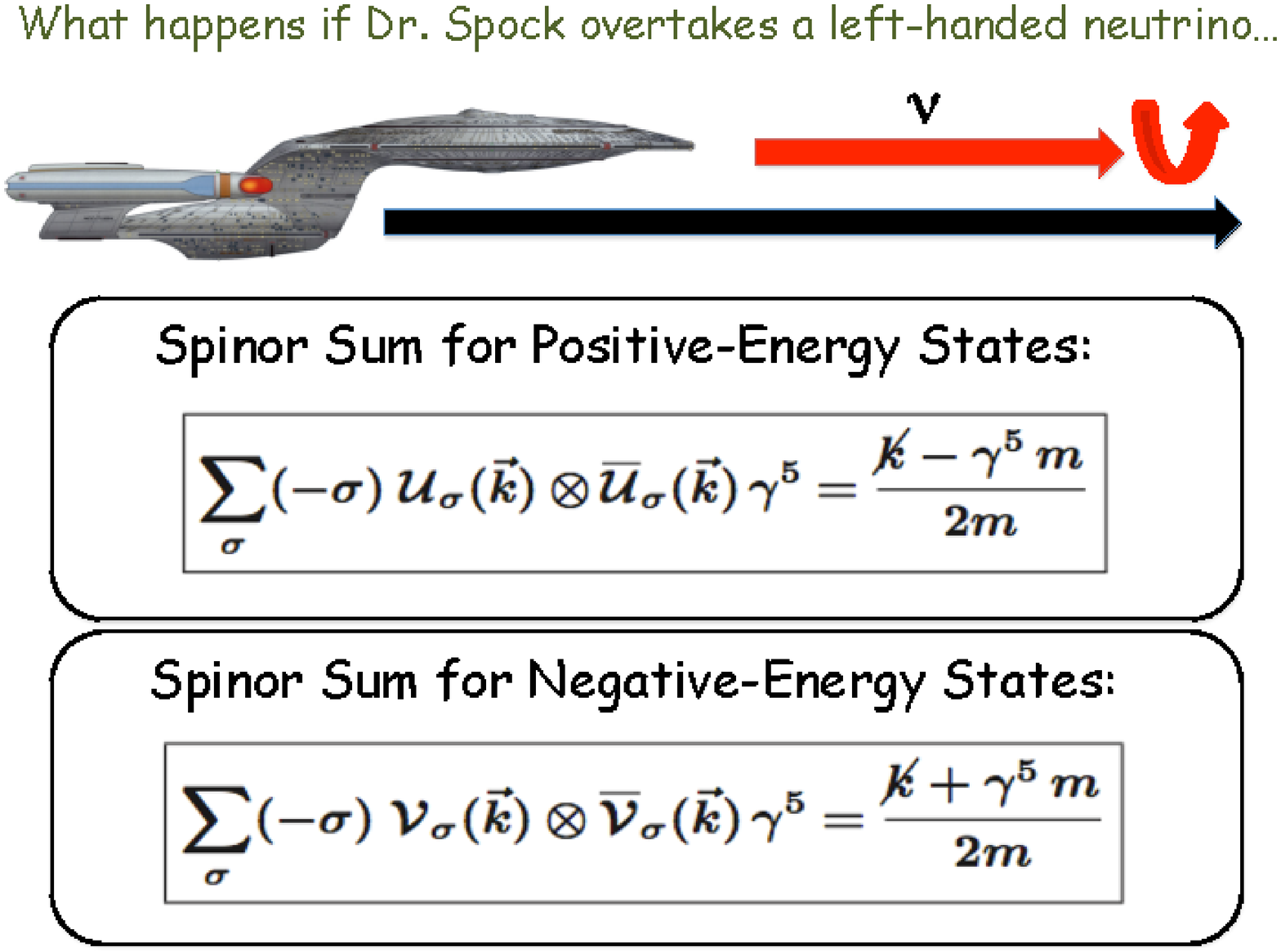}\\
\end{center}
{\em
There is a paradox which results if one overtakes a left-handed
neutrino, looks back and sees the same particle right-handed,
because right-handed neutrinos have never been observed in
nature. One possible way to solve the paradox is sketched
in the current article; the spinor sums are given in Eq.~\eqref{spinorsum5}.
Further explanations are in the text.
}
\end{minipage}
\end{center}

\end{document}